\documentclass[a4paper,11pt]{article}
\usepackage[left=20mm,right=20mm,top=20mm,bottom=20mm]{geometry}
\usepackage{amsmath}
\usepackage{amsfonts}
\usepackage{amssymb}
\usepackage{amsbsy}
\usepackage{color,graphics}
\usepackage{setspace}
\usepackage{epsfig}
\usepackage{cite}
\usepackage{graphicx}
\usepackage{latexsym,multicol}
\newtheorem{theorem}{Theorem}[section]

\newtheorem{definition}[theorem]{Definition}

\def\bproof{\noindent\textbf{Proof: }}
\def\eproof{$\Box$}
\graphicspath{{./SEOsc-figs/}}
\begin{document}
\title{Likely oscillatory motions of stochastic hyperelastic solids}
\author{L. Angela Mihai\footnote{School of Mathematics, Cardiff University, Senghennydd Road, Cardiff, CF24 4AG, UK, Email: \texttt{MihaiLA@cardiff.ac.uk}}
	\quad Danielle Fitt\footnote{School of Mathematics, Cardiff University, Senghennydd Road, Cardiff, CF24 4AG, UK, Email: \texttt{FittD@cardiff.ac.uk}}
	\quad Thomas E. Woolley\footnote{School of Mathematics, Cardiff University, Senghennydd Road, Cardiff, CF24 4AG, UK, Email: \texttt{WoolleyT1@cardiff.ac.uk}}
	\quad Alain Goriely\footnote{Mathematical Institute, University of Oxford, Woodstock Road, Oxford, OX2 6GG, UK, Email: \texttt{goriely@maths.ox.ac.uk}}
}	\date{April 30, 2019}
\maketitle

\begin{abstract}
Stochastic homogeneous hyperelastic solids are characterised by strain-energy densities where the parameters are random variables defined by probability density functions. These models allow for the propagation of uncertainties from input data to output quantities of interest. To investigate the effect of probabilistic parameters on predicted mechanical responses, we study radial oscillations of cylindrical and spherical shells of stochastic incompressible isotropic hyperelastic material, formulated as quasi-equilibrated motions where the system is in equilibrium at every time instant. Additionally, we study finite shear oscillations of a cuboid, which are not quasi-equilibrated. We find that, for hyperelastic bodies of stochastic neo-Hookean or Mooney-Rivlin material, the amplitude and period of the oscillations follow probability distributions that can be characterised. Further, for cylindrical tubes and spherical shells, when an impulse surface traction is applied, there is a parameter interval where the oscillatory and non-oscillatory motions compete, in the sense that both have a chance to occur with a given probability. We refer to the dynamic evolution of these elastic systems, which exhibit inherent uncertainties due to the material properties, as ``likely oscillatory motions''. \\
	
\end{abstract}


\begin{quote}
	\emph{``Denominetur motus talis, qualis omni momento temporis $t$ praebet configurationem capacem aequilibrii corporis iisdem viribus massalibus sollicitati, `motus quasi aequilibratus'. Generatim motus quasi aequilibratus non congruet legibus dynamicis et proinde motus verus corporis fieri non potest, manentibus iisdem viribus masalibus.''} - C. Truesdell (1962) \cite{Truesdell:1962}
\end{quote}

\section{Introduction}

Motivated by numerous long-standing and modern engineering problems, oscillatory motions of cylindrical and spherical shells made of linear elastic material \cite{Love:1888,Love:1944,Krauss:1967,Reissner:1941} have generated a wide range of experimental, theoretical, and computational studies \cite{Alijani:2014:AA,Amabili:2008,Amabili:2003:AP,Breslavsky:2018:BA,Dong:2018:etal}. In contrast, time-dependent finite oscillations of cylindrical tubes and spherical shells of nonlinear hyperelastic material, relevant to the modelling of physical responses in many biological and synthetic systems \cite{Ahamed:2018:etal,Aranda:2017:etal,Destrade:2011:DGS,Haas:2019:HG,Haas:2015:HG,Haslach:2004:HH,Kumar:2013:CDG}, have been less investigated, and much of the work in finite nonlinear elasticity has focused on the static stability of pressurised shells  \cite{Adkins:1952:AR,Biscari:2010:BO,Bucchi:2013a:BH,Bucchi:2013b:BH,Carroll:1987,Fu:2016:FLF,Goncalves:2008:GPL,Goriely:2006:GDBA,Green:1950:GS,Mangan:2015:MD,Muller:2002:MS,Rivlin:1949:VI,Shield:1972,Zamani:2017:ZP}, or on wave-type solutions in infinite media \cite{Ilichev:2014:IF,Pearce:2010:PF}.

The governing equations for large amplitude oscillations of cylindrical tubes and spherical shells of homogeneous isotropic incompressible nonlinear hyperelastic material, formulated as special cases of \emph{quasi-equilibrated motions} \cite{Truesdell:1962}, were reviewed in \cite{TruesdellNoll:2004}. These are the class of motions for which the deformation field is circulation preserving, and at every time instant, the current configuration is a possible  static configuration under the given forces. The free and forced axially symmetric radial oscillations of infinitely long, isotropic incompressible circular cylindrical tubes, with arbitrary wall thickness, were described for the first time in \cite{Knowles:1960,Knowles:1962}. In \cite{Heng:1963:HS,Knowles:1965:KJ,Wang:1965}, free and forced oscillations of spherical shells were derived analogously. For the combined radial-axial large amplitude oscillations of hyperelastic cylindrical tubes, in \cite{Shahinpoor:1973}, the surface tractions necessary to maintain the periodic motions were discussed, and the results were applied to a tube sealed at both ends and filled with an incompressible fluid. The dynamic deformation of cylindrical tubes of Mooney-Rivlin material in finite amplitude radial oscillation was obtained in \cite{Shahinpoor:1971:SN,Shahinpoor:1973,Shahinpoor:1978:SB}. Oscillatory motion caused by the dynamic cavitation of a neo-Hookean sphere was considered in \cite{ChouWang:1989b:CWH}. For a wide class of hyperelastic materials, both the static and dynamic cavitation of homogeneous spheres were analysed in \cite{Ball:1982}. For a hyperelastic sphere of Mooney-Rivlin material, with a cavity, the solution to the nonlinear problem of large amplitude oscillations was computed numerically in \cite{Balakrishnan:1978:BS}. Theoretical and experimental studies of cylindrical and spherical shells of rubberlike material under external pressure were presented in \cite{Wang:1972:WE}. In \cite{Calderer:1983}, the finite amplitude radial oscillations of homogeneous isotropic incompressible hyperelastic spherical and cylindrical shells under a constant pressure difference between the inner and the outer surface were studied theoretically. The finite longitudinal, or `telescopic', oscillations of infinitely long cylindrical tubes were investigated in \cite{Nowinski:1964:NS}. In \cite{Nowinski:1966}, the oscillatory motions of cylindrical and prismatic bodies of incompressible hyperelastic material under dynamic finite shear deformation were analysed. Other dynamic shear deformations were considered in \cite{Wang:1969}, where it was emphasised that such shear motions were not quasi-equilibrated. In \cite{Huilgol:1967}, the dynamic problem of axially symmetric oscillations of cylindrical tubes of transversely isotropic incompressible material, with radial transverse isotropy, was treated. The dynamic deformation of a longitudinally anisotropic thin-walled cylindrical tube under radial oscillations was obtained in \cite{Shahinpoor:1974}. In \cite{Ertepinar:1976:EA}, radial oscillations of non-homogeneous thick-walled cylindrical and spherical shells of neo-Hookean material, with a material constant varying continuously along the radial direction, were studied. In \cite{Akyuz:1998:AE}, for pressurised homogeneous isotropic compressible hyperelastic tubes of arbitrary wall thickness under uniform radial dead-load traction, the stability of the finitely deformed state and small radial vibrations about this state were treated, using the theory of small deformations superposed on large elastic deformations, while the governing equations were solved numerically. In \cite{Verron:1999:VKDR}, the dynamic inflation of hyperelastic spherical membranes of Mooney-Rivlin material subjected to a uniform step pressure was studied, and the absence of damping in these models was discussed. It was concluded that, as the amplitude and period of oscillations are strongly influenced by the rate of internal pressure, if the pressure was suddenly imposed and the inflation process was short, then sustained oscillations due to the dominant elastic effects could be observed. However, for many systems under slowly increasing pressure, strong damping would generally preclude oscillations \cite{DePascalis:etal:2018}. More recently, the dynamic response of incompressible hyperelastic cylindrical and spherical shells subjected to periodic loading was discussed in \cite{Ren:2008,Ren:2009}. Radial oscillations of cylindrical tubes and spherical shells of neo-Hookean \cite{Treloar:1944}, Mooney-Rivlin \cite{Mooney:1940,Rivlin:1948}, and Gent \cite{Gent:1996} hyperelastic materials were analysed in \cite{Beatty:2007,Beatty:2011}, where it was inferred that, in general, both the amplitude and period of oscillations decrease when the stiffness of the material increases. The influence of material constitutive law on the dynamic behaviour of cylindrical and spherical shells was also examined in \cite{Aranda:2015:etal,Aranda:2018:etal,Martinez:2015:etal,Yuan:2008:YZZ}, where the results for Yeoh \cite{Yeoh:1993} and Mooney-Rivlin material models were compared. In \cite{Breslavsky:2016:BAL}, the static and dynamic behaviour of circular cylindrical shells of homogeneous isotropic incompressible hyperelastic material modelling arterial walls were considered. In \cite{Soares:etal:2019}, the nonlinear static and dynamic behaviour of a spherical membrane of neo-Hookean or Mooney-Rivlin material, subjected to a uniformly distributed radial pressure on its inner surface, was studied, and a parametric analysis of the influence of the material constants was presented.

For the assessment and prediction of the mechanical responses of engineered and natural materials, additional challenges arise from the uncertainties in their elastic properties inferred from sparse and approximate observational  data \cite{Ghanem:2017:GHO,Hughes:2010:HH,Kaminski:2018:KL,Oden:2018,Ostroja:2007,Sullivan:2015}. For these materials, deterministic approaches, which are based on average data values, can greatly underestimate or overestimate their properties, and stochastic representations accounting also for data dispersion are needed to significantly improve assessment and predictions. In response to this challenge, stochastic elasticity is a fast developing field that combines nonlinear elasticity and stochastic theories in order to significantly improve model predictions by accounting for uncertainties in the mechanical responses of materials. Within this framework, stochastic hyperelastic materials are advanced phenomenological models described by a strain-energy density where the parameters are characterised by probability density functions, as constructed in \cite{Staber:2015:SG,Staber:2016:SG,Staber:2017:SG,Staber:2018:SG,Staber:2019:SGSMI} and \cite{Mihai:2018:MWG}. These models rely on the notion of entropy (or uncertainty) \cite{Shannon:1948,Soni:2017:SG} and on the maximum entropy principle for a discrete probability distribution \cite{Jaynes:1957a,Jaynes:1957b,Jaynes:2003}, and allow for the propagation of uncertainties from input data to output quantities of interest \cite{Soize:2013}. They are also suitable for incorporation into Bayesian methodologies \cite{Bayes:1763,McGrayne:2012}) for models selection or updates \cite{Mihai:2018:MWG,Oden:2018,Robert:2007}.

To study the effect of probabilistic model parameters on predicted mechanical responses, in \cite{Mihai:2018a:MDWG,Mihai:2018b:MDWG,Mihai:2019a:MWG,Mihai:2019b:MWG}, for different bodies with simple geometries at finite strain deformations, it was shown explicitly that, in contrast to the deterministic elastic problem where a single critical value strictly separates the stable and unstable cases, for the stochastic problem, there is a probabilistic interval where the stable and unstable states always compete, in the sense that both have a quantifiable chance to be found. In addition, revisiting these problems from a novel perspective offered fresh opportunities for gaining new insights into the fundamental elastic solutions, and correcting some inconsistencies found in the previous works. Specific case studies, so far, include the cavitation of a sphere under uniform tensile dead load \cite{Mihai:2018a:MDWG}, the inflation of pressurised spherical and cylindrical shells \cite{Mihai:2018b:MDWG}, the classical problems of the Rivlin cube \cite{Mihai:2019a:MWG}, and the rotation and perversion of anisotropic hyperelastic cylindrical tubes \cite{Mihai:2019b:MWG}.

In this paper, we extend the stochastic framework developed in \cite{Mihai:2018a:MDWG,Mihai:2018b:MDWG,Mihai:2019a:MWG,Mihai:2019b:MWG} to study radial oscillations of cylindrical and spherical shells of stochastic incompressible isotropic hyperelastic material formulated as quasi-equilibrated motions. For these motions, the system is in equilibrium at every time instant. We consider also finite shear oscillations of a cuboid, which are not quasi-equilibrated. We find that, for hyperelastic bodies of stochastic neo-Hookean or Mooney-Rivlin material, the amplitude and period of the oscillations follow probability distributions that can be fully characterised. Further, for cylindrical tubes and spherical shells, when an impulse surface traction is applied, there is a parameter interval where the oscillatory and non-oscillatory motions compete in the sense that both have a chance to occur with a given probability. We refer to the dynamic evolution of these elastic systems, which exhibit inherent uncertainties due to the material properties, as ``likely oscillatory motions''.  Section~\ref{sec:prerequisites} provides a summary of the stochastic elasticity prerequisites. Section~\ref{sec:cubes} is devoted to the oscillatory motions of a stochastic hyperelastic cuboid under dynamic generalised shear. This is followed, in Sections~\ref{sec:tube} and \ref{sec:sphere}, by the radial oscillatory motions of stochastic cylindrical and spherical shells with bounded wall thickness, respectively. The limiting cases of thin- and infinitely thick-walled structures are also discussed. Some less straight-forward calculations, inherent for these problems, are deferred to Appendix~\ref{sec:append}. Concluding remarks are drawn in Section~\ref{sec:conclude}.

\section{Prerequisites}\label{sec:prerequisites}

In this section, we recall the notion of (universal) quasi-equilibrated motion in finite elasticity, introduced in \cite{Truesdell:1962} and reviewed in \cite{TruesdellNoll:2004}, and summarise the stochastic finite elasticity framework developed in \cite{Mihai:2018:MWG} and applied to various static stability problems in \cite{Mihai:2018a:MDWG,Mihai:2018b:MDWG,Mihai:2019a:MWG}.

\subsection{Quasi-equilibrated motion}

For the large strain time-dependent behaviour of an elastic solid, Cauchy's laws of motion (balance laws of linear and angular momentum) are governed by the following Eulerian field equations \cite[p.~40]{TruesdellNoll:2004},
\begin{eqnarray}
&&\rho\ddot{\textbf{x}}=\mathrm{div}\ \textbf{T}+\rho\textbf{b},\label{eq:1st}\\
&&\textbf{T}=\textbf{T}^{T},\label{eq:2nd}
\end{eqnarray}
where $\textbf{x}=\chi(\textbf{X},t)$ is the motion of the elastic solid, $\rho$ is the material density, which is assumed constant, $\textbf{b}=\textbf{b}(\textbf{x},t)$ is the body force, $\textbf{T}=\textbf{T}(\textbf{x},t)$ is the Cauchy stress tensor, and the superscript $T$ defines the transpose. To obtain possible dynamical solutions, one can solve Cauchy's equation for particular motions, or generalise known static solutions to dynamical forms, using the so-called \textit{quasi-equilibrated motion}, which is defined as follows.

\begin{definition}\label{def:qem} \cite[p.~208]{TruesdellNoll:2004}
	A quasi-equilibrated motion, $\textbf{x}=\chi(\textbf{X},t)$, is the motion of an incompressible homogeneous elastic solid subject to a given body force, $\textbf{b}=\textbf{b}(\textbf{x},t)$, whereby, for each value of $t$, $\textbf{x}=\chi(\textbf{X},t)$ defines a static deformation that satisfies the equilibrium conditions under the body force $\textbf{b}=\textbf{b}(\textbf{x},t)$.
\end{definition}

\begin{theorem}\label{th:qem} \cite[p.~208]{TruesdellNoll:2004}
	A quasi-equilibrated motion, $\textbf{x}=\chi(\textbf{X},t)$, of an incompressible homogeneous elastic solid subject to a given body force, $\textbf{b}=\textbf{b}(\textbf{x},t)$, is dynamically possible, subject to the same body force, if and only if the motion is circulation preserving with a single-valued acceleration potential $\xi$, i.e.,
	\begin{equation}\label{eq:cp}
	\ddot{\textbf{x}}=-\mathrm{grad}\ \xi.
	\end{equation}
	For the condition \eqref{eq:cp} to be satisfied, it is necessary that
	\begin{equation}\label{eq:ddotx:curl}
	\mathrm{curl}\ \ddot{\textbf{x}}=\textbf{0}.
	\end{equation}
	Then, the Cauchy stress tensor takes the form
	\begin{equation}\label{eq:T}
	\textbf{T}=-\rho\xi\textbf{I}+\textbf{T}^{(0)},
	\end{equation}
	where $\textbf{T}^{(0)}$ is the Cauchy stress for the equilibrium state at time $t$ and $\textbf{I}=\text{diag}(1,1,1)$ is the identity tensor. In this case, the stress field is determined by the present configuration alone. In particular, the shear stresses in the motion are the same as those of the equilibrium state at time $t$.
\end{theorem}

\bproof
The Cauchy stress $\textbf{T}^{(0)}$ for the equilibrium state under the body force $\textbf{b}=\textbf{b}(\textbf{x},t)$ at time $t$ satisfies 
\begin{equation}\label{eq:1:proof}
-\mathrm{div}\ \textbf{T}^{(0)}=\rho\textbf{b}.
\end{equation}
First, we assume that the motion $\textbf{x}=\chi(\textbf{X},t)$ is quasi-equilibrated under the body force $\textbf{b}=\textbf{b}(\textbf{x},t)$, and deduce that there is a single-valued function $\xi$, such that \eqref{eq:cp} holds. Indeed, if the motion is quasi-equilibrated, then  Definition~\ref{def:qem} implies that, at any fixed time-instant $t$, the Cauchy stress takes the form \eqref{eq:T}, where $\xi=\xi(t)$ is a single-valued function of $t$. Substituting \eqref{eq:T} in \eqref{eq:1st} gives
\begin{equation}\label{eq:2:proof}
\rho\ddot{\textbf{x}}=-\rho\ \mathrm{grad}\ \xi+\mathrm{div}\ \textbf{T}^{(0)}+\rho\textbf{b}.
\end{equation}
Then, \eqref{eq:cp} follows from \eqref{eq:1:proof} and \eqref{eq:2:proof}.

Conversely, if \eqref{eq:cp} holds, with $\xi$ a single-valued function, then, substitution of \eqref{eq:cp} and \eqref{eq:1:proof} in \eqref{eq:1st} gives
\begin{equation}\label{eq:3:proof}
-\rho\ \mathrm{grad}\ \xi=\mathrm{div}\ \left(\textbf{T}-\textbf{T}^{(0)}\right),
\end{equation}
at any time-instant $t$. From \eqref{eq:3:proof}, it follows that the Cauchy stress $\textbf{T}$ takes the form \eqref{eq:T}. Hence, the motion is quasi-equilibrated according to  Definition~\ref{def:qem}. \eproof\\

Theorem~\ref{th:qem} may only be applicable to specific quasi-equilibrated motions of specific materials. Nevertheless, by the above theorem, for a quasi-equilibrated motion to be dynamically possible under a given body force in all elastic materials, it is necessary that, at every time instant, the deformation is a possible equilibrium state under that body force in all those materials. Quasi-equilibrated motions of isotropic materials subject to surface tractions alone are obtained by taking the arbitrary constant in those deformations to be arbitrary functions of time. Some examples are the homogeneous motions that are possible in all homogeneous incompressible materials, and also those considered by us in Sections~\ref{sec:tube} and \ref{sec:sphere}  (see also \cite[p.~209]{TruesdellNoll:2004}).

\subsection{Stochastic isotropic incompressible hyperelastic models}

A stochastic homogeneous hyperelastic model is defined by a stochastic strain-energy function, for which the model parameters are random variables, drawn from probability distributions \cite{Mihai:2018:MWG,Staber:2015:SG,Staber:2016:SG,Staber:2017:SG}. In this case, each model parameter is usually described in terms of its \emph{mean value} and its \emph{variance}, which contains information about the range of values about the mean value \cite{Brewick:2018:BT,Caylak:2018:etal,Hughes:2010:HH,McCoy:1973,Norenberg:2015:NM}. Here, we combine finite elasticity  \cite{goriely17,Ogden:1997,TruesdellNoll:2004} and probability theory \cite{Grimmett:2001:GS,Jaynes:2003}, and rely on the following general assumptions \cite{Mihai:2018a:MDWG,Mihai:2018b:MDWG,Mihai:2018:MWG,Mihai:2019a:MWG}:
\begin{itemize}
	\item[(A1)] Material objectivity, stating that constitutive equations must be invariant under changes of frame of reference. This requires that the scalar strain-energy function, $W=W(\textbf{F})$, depending only on the deformation gradient $\textbf{F}$, with respect to the reference configuration, is unaffected by a superimposed rigid-body transformation (which involves a change of position) after deformation, i.e., $W(\textbf{R}^{T}\textbf{F})=W(\textbf{F})$, where $\textbf{R}\in SO(3)$ is a proper orthogonal tensor (rotation). Material objectivity is guaranteed by defining strain-energy functions in terms of invariants.
	
	\item[(A2)] Material isotropy, requiring that the strain-energy function is unaffected by a superimposed rigid-body transformation prior to deformation, i.e., $W(\textbf{F}\textbf{Q})=W(\textbf{F})$, where $\textbf{Q}\in SO(3)$. For isotropic materials, the  strain-energy  function is a symmetric function of the  principal stretches $\{\lambda_{i}\}_{i=1,2,3}$ of $\textbf{F}$, i.e., $W(\textbf{F})=\mathcal{W}(\lambda_{1},\lambda_{2},\lambda_{3})$.
	
	\item[(A3)] Baker-Ericksen (BE) inequalities, which state that \emph{the greater principal (Cauchy) stress occurs in the direction of the greater principal stretch}, are \cite{BakerEricksen:1954}
	\begin{equation}\label{eq:BE}
	\left(T_{i}-T_{j}\right)\left(\lambda_{i}-\lambda_{j}\right)>0\quad \mbox{if}\quad \lambda_{i}\neq\lambda_{j},\quad i,j=1,2,3,
	\end{equation}
	where $\{\lambda_{i}\}_{i=1,2,3}$ and $\{T_{i}\}_{i=1,2,3}$ denote the principal stretches and the principal Cauchy stresses, respectively.  The BE inequalities \eqref{eq:BE} take the equivalent form
	\begin{equation}\label{eq:W:BE}
	\left(\lambda_{i}\frac{\partial\mathcal{W}}{\partial\lambda_{i}}-\lambda_{j}\frac{\partial\mathcal{W}}{\partial\lambda_{j}}\right)\left(\lambda_{i}-\lambda_{j}\right)>0\quad \mbox{if}\quad \lambda_{i}\neq\lambda_{j},\quad i,j=1,2,3.
	\end{equation}
	In \eqref{eq:BE}-\eqref{eq:W:BE}, the strict inequality ``$>$'' is replaced by ``$\geq$'' if any two principal stretches are equal.

	\item[(A4)] Finite mean and variance of the random shear modulus, i.e., at any given deformation, the random shear modulus, $\mu$, and its inverse, $1/\mu$, are second-order random variables \cite{Staber:2015:SG,Staber:2016:SG,Staber:2017:SG}.
\end{itemize}
Assumptions (A1)-(A3) are well-known principles in isotropic finite elasticity \cite{goriely17,Ogden:1997,TruesdellNoll:2004}. In particular, regarding (A3), we recall that, for a homogeneous hyperelastic body under uniaxial tension, the deformation is a simple extension in the direction of the tensile force if and only if the BE inequalities \eqref{eq:BE} hold \cite{Marzano:1983}. Another important deformation is that of simple shear superposed on axial stretch, defined by
\begin{equation}\label{eq:stretchshear}
x_1=\alpha X_1+k\frac{X_2}{\alpha^2},\qquad x_2=\frac{X_2}{\alpha^2},\qquad x_3=\alpha X_3,
\end{equation}
where $(X_1,X_2,X_3)$ and $(x_1,x_2,x_3)$ are the Cartesian coordinates for the reference (Lagrangian) and the current (Eulerian) configuration, respectively, and $k>0$ and $\alpha>0$ are positive constants representing the shear parameter and the axial stretch ($0<\alpha<1$ for axial tension and $\alpha>1$ for axial compression), respectively. For this deformation, the principal stretches, $\{\lambda_{i}\}_{i=1,2,3}$, satisfy
\begin{equation}\label{eq:stretchshear:lambdai}
\begin{split}
\lambda_{1}^2&=\frac{\alpha^6+k^2+1+\sqrt{\left(\alpha^6+k^2+1\right)^2-4\alpha^6}}{2\alpha^4},
\\
\lambda_{2}^2&=\frac{\alpha^6+k^2+1-\sqrt{\left(\alpha^6+k^2+1\right)^2-4\alpha^6}}{2\alpha^4},
\\
\lambda_{3}^2&=\alpha^2.
\end{split}
\end{equation}
Then, assuming that the material is incompressible, the associated principal Cauchy stresses take the form
\begin{equation}\label{eq:sigmai}
T_{i}=\lambda_{i}\frac{\partial\mathcal{W}}{\partial{\lambda_{i}}}-p,\qquad i=1,2,3,
\end{equation}
where $p$ is the Lagrange multiplier for the incompressibility constraint. In this case, if the BE inequalities \eqref{eq:BE} hold, then the nonlinear shear modulus defined by \cite{Mihai:2017:MG,Mihai:2018:MWG}
\begin{equation}\label{eq:nlshearmod}
\widetilde{\mu}=\frac{T_{1}-T_{2}}{\lambda_{1}^2-\lambda_{2}^2}
\end{equation}
is positive, i.e., $\widetilde{\mu}>0$, for all $k>0$ and $\alpha>0$. In the linear elastic limit, where $k\to0$ and $\alpha\to 1$, the nonlinear shear modulus given by \eqref{eq:nlshearmod} converges to the classical shear modulus under infinitesimal deformation, i.e.,
\begin{equation}\label{eq:shearmod}
\lim_{a\to1,k\to0}\widetilde{\mu}=\mu.
\end{equation}
Assumption (A4) then contains physically realistic expectations on the (positive) random shear modulus $\mu>0$, which will be characterised by a suitable probability density function.

In the next sections, we analyse the dynamic generalised shear deformation of a cuboid and the radially symmetric motion of cylindrical tube and spherical shells of stochastic isotropic incompressible hyperelastic material. One can regard a stochastic hyperelastic body as an ensemble of bodies with the same geometry, where each individual body is made from a homogeneous isotropic incompressible hyperelastic material, with the elastic parameters drawn from probability distributions. Then, for the individual hyperelastic bodies, the finite elasticity theory applies.

Throughout this paper, we confine our attention to a class of stochastic homogeneous incompressible hyperelastic materials described by the Mooney-Rivlin-like constitutive law \cite{Mihai:2018:MWG,Staber:2015:SG},
\begin{equation}\label{eq:W:stoch}
\mathcal{W}(\lambda_{1},\lambda_{2},\lambda_{3})=\frac{\mu_{1}}{2}\left(\lambda_{1}^2+\lambda_{2}^2+\lambda_{3}^2-3\right)
+\frac{\mu_{2}}{2}\left(\lambda_{1}^{-2}+\lambda_{2}^{-2}+\lambda_{3}^{-2}-3\right),
\end{equation}
where $\mu_{1}$ and $\mu_{2}$ are random variables. The non-deterministic model \eqref{eq:W:stoch} reduces to a stochastic neo-Hookean model if $\mu_{2}=0$. If the random parameters $\mu_{1}$ and $\mu_{2}$ are replaced by their respective mean values, $\underline{\mu}_{1}$ and $\underline{\mu}_{2}$, then the resulting mean hyperelastic model coincides with the usual deterministic one.

For the stochastic material model described by \eqref{eq:W:stoch}, the shear modulus in infinitesimal deformation is defined as $\mu=\mu_{1}+\mu_{2}$. For this modulus, we set the following mathematical constraints, which ensure that the assumption (A4), made in Section~\ref{sec:prerequisites} is satisfied \cite{Mihai:2018:MWG},
\begin{eqnarray}\label{eq:Emu1}\begin{cases}
E\left[\mu\right]=\underline{\mu}>0,&\\
E\left[\log\ \mu\right]=\nu,& \mbox{such that $|\nu|<+\infty$},\label{eq:Emu2}\end{cases}
\end{eqnarray}
i.e., the mean value $\underline{\mu}$ of the shear modulus $\mu$ is fixed and greater than zero, and the mean value of $\log\ \mu$ is fixed and finite. It follows that $\mu$ and $1/\mu$ are second-order random variables, i.e., they have finite mean and finite variance \cite{Soize:2000,Soize:2001}. Under the constraints (\ref{eq:Emu1}), $\mu$ follows a Gamma probability distribution with hyperparameters $\rho_{1}>0$ and $\rho_{2}>0$, such that
\begin{equation}\label{eq:rho12}
\underline{\mu}=\rho_{1}\rho_{2},\qquad
\text{Var}[\mu]=\rho_{1}\rho_{2}^2,
\end{equation}
where $\underline{\mu}$ is the mean value, $\text{Var}[\mu]$ is the variance , and $\|\mu\|=\sqrt{\text{Var}[\mu]}$ is the standard deviation of $\mu$. The corresponding probability density function takes the form \cite{Abramowitz:1964,Johnson:1994:JKB}
\begin{equation}\label{eq:mu:gamma}
g(\mu;\rho_{1},\rho_{2})=\frac{\mu^{\rho_{1}-1}e^{-\mu/\rho_{2}}}{\rho_{2}^{\rho_{1}}\Gamma(\rho_{1})},\qquad\mbox{for}\ \mu>0\ \mbox{and}\ \rho_{1}, \rho_{2}>0,
\end{equation}
where $\Gamma:\mathbb{R}^{*}_{+}\to\mathbb{R}$ is the complete Gamma function
\begin{equation}\label{eq:gamma}
\Gamma(z)=\int_{0}^{+\infty}t^{z-1}e^{-t}\text dt.
\end{equation}
When $\mu_{1}>0$ and $\mu_{2}>0$, we can define the auxiliary random variable \cite{Mihai:2018:MWG}
\begin{equation}\label{eq:R12:b}
R_{1}=\frac{\mu_{1}}{\mu},
\end{equation}
such that $0<R_{1}<1$. Then
\begin{equation}\label{eq:mu12:b}
\mu_{1}=\mu R_{1},\qquad \mu_{2}=\mu-\mu_{1}=\mu(1-R_{1}).
\end{equation}
Setting the realistic constraints \cite{Mihai:2018:MWG,Staber:2015:SG,Staber:2016:SG,Staber:2017:SG},
\begin{eqnarray}\begin{cases}
E\left[\log\ R_{1}\right]=\nu_{1},& \mbox{such that $|\nu_{1}|<+\infty$},\label{eq:ER1}\\
E\left[\log(1-R_{1})\right]=\nu_{2},& \mbox{such that $|\nu_{2}|<+\infty$},\label{eq:ER2}\end{cases}
\end{eqnarray}
we obtain that $R_{1}$ follows a standard Beta distribution \cite{Abramowitz:1964,Johnson:1994:JKB}, with hyperparameters $\xi_{1}>0$ and $\xi_{2}>0$ satisfying
\begin{equation}\label{eq:xi12}
\underline{R}_{1}=\frac{\xi_{1}}{\xi_{1}+\xi_{2}},\qquad
\text{Var}[R_{1}]=\frac{\xi_{1}\xi_{2}}{\left(\xi_{1}+\xi_{2}\right)^2\left(\xi_{1}+\xi_{2}+1\right)}.
\end{equation}
where $\underline{R}_{1}$ is the mean value, $\text{Var}[R_{1}]$ is the variance, and $\|R_{1}\|=\sqrt{\text{Var}[R_{1}]}$ is the standard deviation of $R_{1}$. The corresponding probability density function takes the form
\begin{equation}\label{eq:betar}
\beta(r;\xi_{1},\xi_{2})=\frac{r^{\xi_{1}-1}(1-r)^{\xi_{2}-1}}{B(\xi_{1},\xi_{2})},\qquad \qquad\mbox{for}\ r\in(0,1)\ \mbox{and}\ \xi_{1}, \xi_{2}>0,
\end{equation}
where $B:\mathbb{R}^{*}_{+}\times\mathbb{R}^{*}_{+}\to\mathbb{R}$ is the Beta function
\begin{equation}\label{eq:beta}
B(x,y)=\int_{0}^{1}t^{x-1}(1-t)^{y-1}dt.
\end{equation}
Thus, for the random coefficients given by \eqref{eq:mu12:b}, the corresponding mean values take the form,
\begin{equation}\label{eq:mu12:mean}
\underline{\mu}_{1}=\underline{\mu}\underline{R}_{1},
\qquad \underline{\mu}_{2}=\underline{\mu}-\underline{\mu}_{1}=\underline{\mu}(1-\underline{R}_{1}),
\end{equation}
and the variances and covariance are, respectively,
\begin{eqnarray}
&&\text{Var}\left[\mu_{1}\right]
=(\underline{\mu})^2\text{Var}[R_{1}]+(\underline{R}_{1})^2\text{Var}[\mu]+\text{Var}[\mu]\text{Var}[R_{1}],\\
&&\text{Var}\left[\mu_{2}\right]
=(\underline{\mu})^2\text{Var}[R_{1}]+(1-\underline{R}_{1})^2\text{Var}[\mu]+\text{Var}[\mu]\text{Var}[R_{1}],\\
&&\text{Cov}[\mu_{1},\mu_{2}]=\frac{1}{2}\left(\text{Var}[\mu]-\text{Var}[\mu_{1}]-\text{Var}[\mu_{2}]\right).
\end{eqnarray}
Note that the random variables $\mu$ and $R_1$ are independent, depending on parameters $(\rho_1,\rho_2)$ and $(\zeta_1,\zeta_2)$, respectively, whereas $\mu_1$ and $\mu_2$ are codependent variables as they both require $(\mu,R_1)$ to be defined. 

\begin{figure}[htbp]
	\begin{center}
		\includegraphics[width=0.45\textwidth]{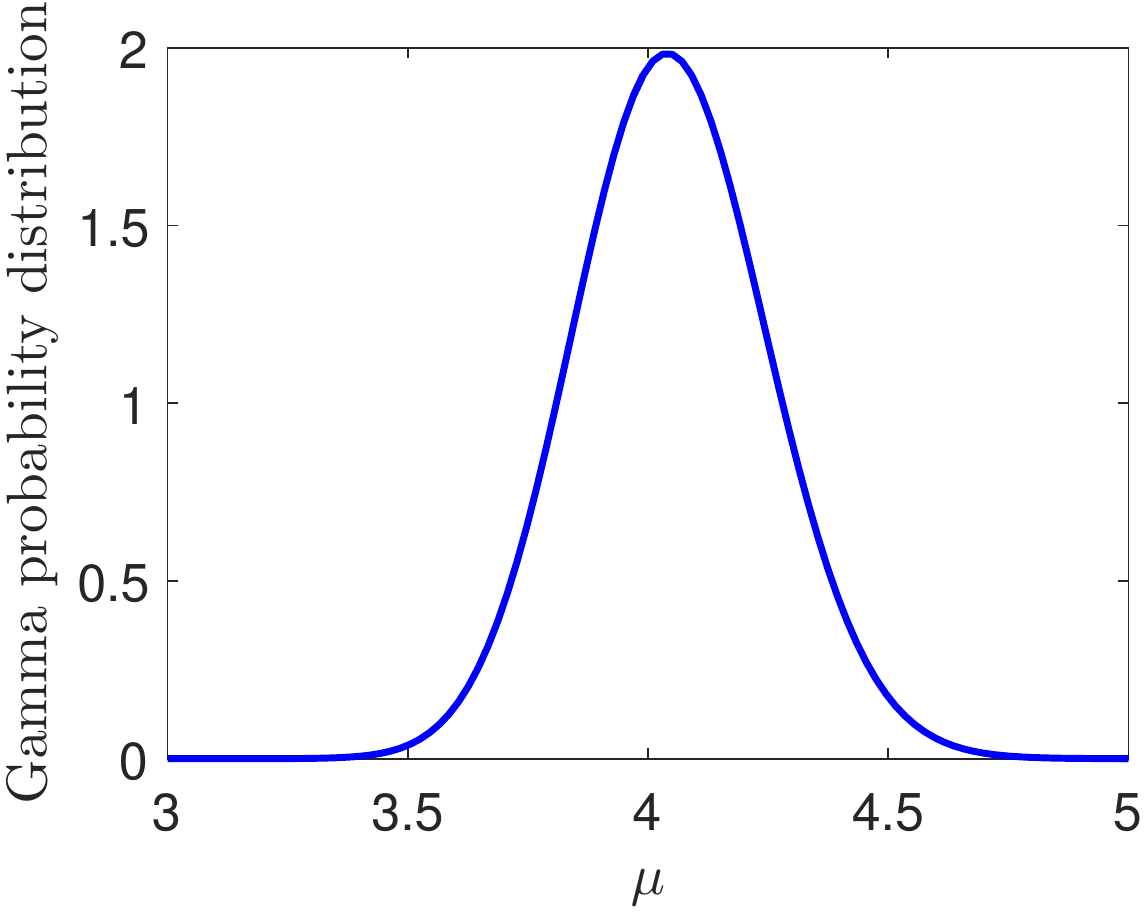}
		\caption{Example of Gamma distribution with hyperparameters $\rho_{1}=405$ and $\rho_{2}=0.01$.}\label{fig:mu-gpdf}
	\end{center}
\end{figure}

For the numerical illustration of our subsequent results, throughout this paper, we assume that the random shear modulus $\mu$ follows the Gamma distribution represented in Figure~\ref{fig:mu-gpdf}, where the shape and scale parameters are $\rho_{1}=405$ and $\rho_{2}=0.01$, respectively \cite{Mihai:2018b:MDWG}. Different simulations were then created by fixing the parameters (given in each figure caption), and repeatedly drawing random samples from the underlying distribution. Our computer simulations were run in Matlab 2018a, where we made specific use of inbuilt functions for random number generation. 
	
Note also that the Gamma distribution represented in Figure~\ref{fig:mu-gpdf}, for which $\rho_1$ is large compared to $\rho_2$, appears to be approximately a normal distribution. However, despite known convergence results and the qualitative agreement between the two density functions for large values of the mean (see \cite{Mihai:2018b:MDWG} for a detailed discussion), in general, the normal distribution cannot be used to model elastic parameters. This is due to the fact that the normal distribution is defined on the entire real line, whereas elastic moduli are typically positive. In practice, these moduli can meaningfully take on different values, corresponding to possible outcomes of the experiments. Then, the maximum entropy principle allows for the explicit construction of their probability laws, given the available information. Explicit derivations of probability distributions for the constitutive parameters of stochastic homogeneous isotropic hyperelastic models, calibrated to experimental data, are presented in \cite{Mihai:2018:MWG,Staber:2017:SG}.

\section{Generalised shear motion of a stochastic hyperelastic cuboid}\label{sec:cubes}

First, we consider a stochastic hyperelastic cuboid subject to dynamic generalised shear.

\subsection{Dynamic generalised shear}

\begin{figure}[htbp]
	\begin{center}
		\includegraphics[width=\textwidth]{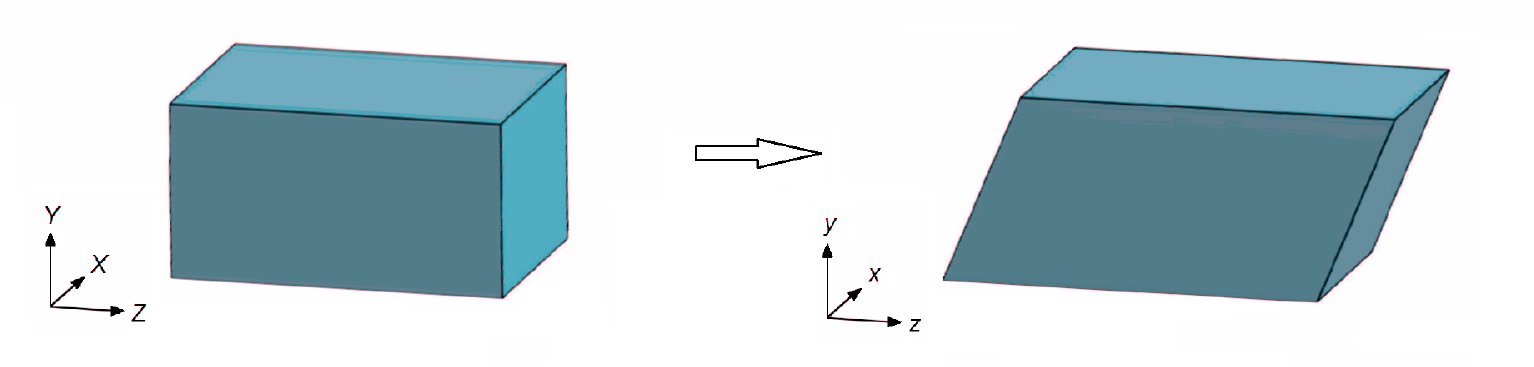}
		\caption{Schematic of generalised shear of a cuboid, showing the reference state (left) and the deformed state (right), respectively.}\label{fig:cube}
	\end{center}
\end{figure}

The generalised shear motion of an elastic body is described by \cite{Destrade:2011:DGS}
\begin{equation}\label{eq:cube:gshear}
x=\frac{X}{\sqrt{\alpha}},\qquad
y=\frac{Y}{\sqrt{\alpha}},\qquad
z=\alpha Z+u\left(X,Y ,t\right),
\end{equation}
where $(X,Y,Z)$ and $(x,y,z)$ are the Cartesian coordinates for the reference (Lagrangian, material) and current (Eulerian, spatial) configuration, respectively, $\alpha>0$ is a given constant, and $u=z-Z$, representing the displacement in the third direction, is a time-dependent function to be determined. Here, we assume that the edges of the cuboid are aligned with the directions of the Cartesian axes in the undeformed state (see Figure~\ref{fig:cube}).

By the governing equations \eqref{eq:cube:gshear}, the condition \eqref{eq:ddotx:curl} is valid for $\textbf{x}=(x,y,z)^{T}$ if and only if
\begin{equation}\label{eq:cube:curl}
\textbf{0}=\mathrm{curl}\ \ddot{\textbf{x}}=
\left[
\begin{array}{c}
\partial\ddot{z}/\partial y-\partial\ddot{y}/\partial z\\
\partial\ddot{x}/\partial z-\partial\ddot{z}/\partial x\\
\partial\ddot{y}/\partial x-\partial\ddot{x}/\partial y
\end{array}
\right]=
\left[
\begin{array}{c}
\partial\ddot{u}/\partial Y\\
-\partial\ddot{u}/\partial X\\
0
\end{array}
\right].
\end{equation}
This condition imposes very strict constraints on the motion. Yet, we will see that even though the generalised shear motion \eqref{eq:cube:gshear} is not quasi-equilibrated, exact solutions are still available, although these solutions are not universal \cite{Nowinski:1966,Wang:1969}.

For the deformation (\ref{eq:cube:gshear}), the gradient tensor is equal to
\[
\textbf{F}
=\left[
\begin{array}{ccc}
1/\sqrt{\alpha} & 0 & 0\\
0 & 1/\sqrt{\alpha} & 0\\
u_{X} & u_{Y} & \alpha
\end{array}
\right],
\]
where $u_{X}$ and $u_{Y}$ denote the partial first derivatives of $u$ with respect to $X$ and $Y$, respectively. The corresponding left Cauchy-Green tensor is
\begin{equation}\label{eq:cube:B}
\textbf{B}=\textbf{F}\textbf{F}^{T}=\left[
\begin{array}{ccc}
1/\alpha & 0 & u_{X}/\sqrt{\alpha} \\
0 & 1/\alpha & u_{Y}/\sqrt{\alpha} \\
u_{X}/\sqrt{\alpha} & u_{Y}/\sqrt{\alpha} & u_{X}^2+ u_{Y}^2+\alpha^2
\end{array}
\right],
\end{equation}
and has the principal invariants
\begin{equation}\label{eq:cube:I123}
\begin{split}
I_{1}=&\mathrm{tr}\ (\textbf{B})=u_{X}^2+u_{Y}^2+\frac{2}{\alpha}+\alpha^2,\\
I_{2}=&\frac{1}{2}\left[\left(\mathrm{tr}\,\textbf{B}\right)^{2}-\mathrm{tr}\left(\textbf{B}^{2}\right)\right]=\frac{u_{X}^2}{\alpha}+\frac{u_{Y}^2}{\alpha}+\frac{1}{\alpha^2}+2\alpha,\\
I_{3}=&\det\textbf{B}=1.
\end{split}
\end{equation}
The associated Cauchy stress tensor takes the form \cite[pp.~87-91]{Green:1970:GA}
\begin{equation}\label{eq:cube:T}
\textbf{T}=-p\textbf{I}+\beta_{1}\textbf{B}+\beta_{-1}\textbf{B}^{-1},
\end{equation}
where $p$ is the Lagrange multiplier for the incompressibility constraint ($I_{3}=1$), and
\begin{equation}\label{eq:cube:betas}
\beta_{1}={2}\frac{\partial W}{\partial I_{1}},\qquad \beta_{-1}=-2\frac{\partial W}{\partial I_{2}}
\end{equation}
are the nonlinear material parameters, with $I_{1}$, $I_{2}$ given by \eqref{eq:cube:I123}.

\subsection{Shear oscillations of a cuboid of stochastic neo-Hookean material}

We now specialise to the case of a cuboid of stochastic neo-Hookean material, with $\mu_{1}=\mu>0$ and $\mu_{2}=0$ in \eqref{eq:W:stoch}, where the non-zero components of the Cauchy stress tensor given by \eqref{eq:cube:T} are as follows
\begin{equation}\label{eq:cube:stresses}
\begin{split}
T_{xx}&=T_{yy}=-p+\frac{\mu}{\alpha},\\
T_{zz}&=-p+\mu\left(u_{X}^2+u_{Y}^2+\alpha^2\right),\\
T_{xz}&=\frac{\mu}{\sqrt{\alpha}} u_{X},\\
T_{yz}&=\frac{\mu}{\sqrt{\alpha}} u_{Y}.
\end{split}
\end{equation}
Then, by the equation of motion \eqref{eq:1st},
\begin{equation}\label{eq:cube:1st}
\begin{split}
&\frac{\partial p}{\partial x}=0,\\
&\frac{\partial p}{\partial y}=0,\\
&\frac{\partial p}{\partial z}=-\rho\ddot{u}+\mu\left(u_{XX}+u_{YY}\right),
\end{split}
\end{equation}
where $u_{XX}$ and $u_{YY}$ represent the second derivatives of $u$ with respect to $X$ and $Y$, respectively. Hence, $p$ is independent of $x$ and $y$.

We consider the undeformed cuboid to be long in the $Z$-direction, and impose an initial displacement $u_{0}(X,Y)=u(X,Y,0)$ and velocity $\dot{u}_{0}(X,Y)=\dot{u}(X,Y,0)$. For the boundary condition, we distinguish the following two cases:
\paragraph{(i)} If we impose null normal Cauchy stresses, $T_{xx}=T_{yy}=0$, on the faces perpendicular to the $X$- and $Y$-directions, at all time, then $p=\mu/\alpha$ is constant and $T_{zz}=\mu\left(u_{X}^2+u_{Y}^2+\alpha^2-1/\alpha\right)$.

\paragraph{{(ii)}} If $T_{xx}=T_{yy}\neq0$, as $T_{zz}$ cannot be made point-wise zero, we denote the normal force acting on the cross-sections of area $A$ in the $z$-direction at time $t$ by
\begin{equation}\label{eq:cube:zforce}
N_{z}(t)=\int_{A} T_{zz}dA,
\end{equation}
and consider this force to be zero, i.e., $N_{z}(t)=0$ at all time. Then, $p$ is independent of $z$, and, by \eqref{eq:cube:1st}, it is also independent of $x$ and $y$, hence, $p=p(t)$.

\begin{figure}[htbp]
	\begin{center}
		\includegraphics[width=\textwidth]{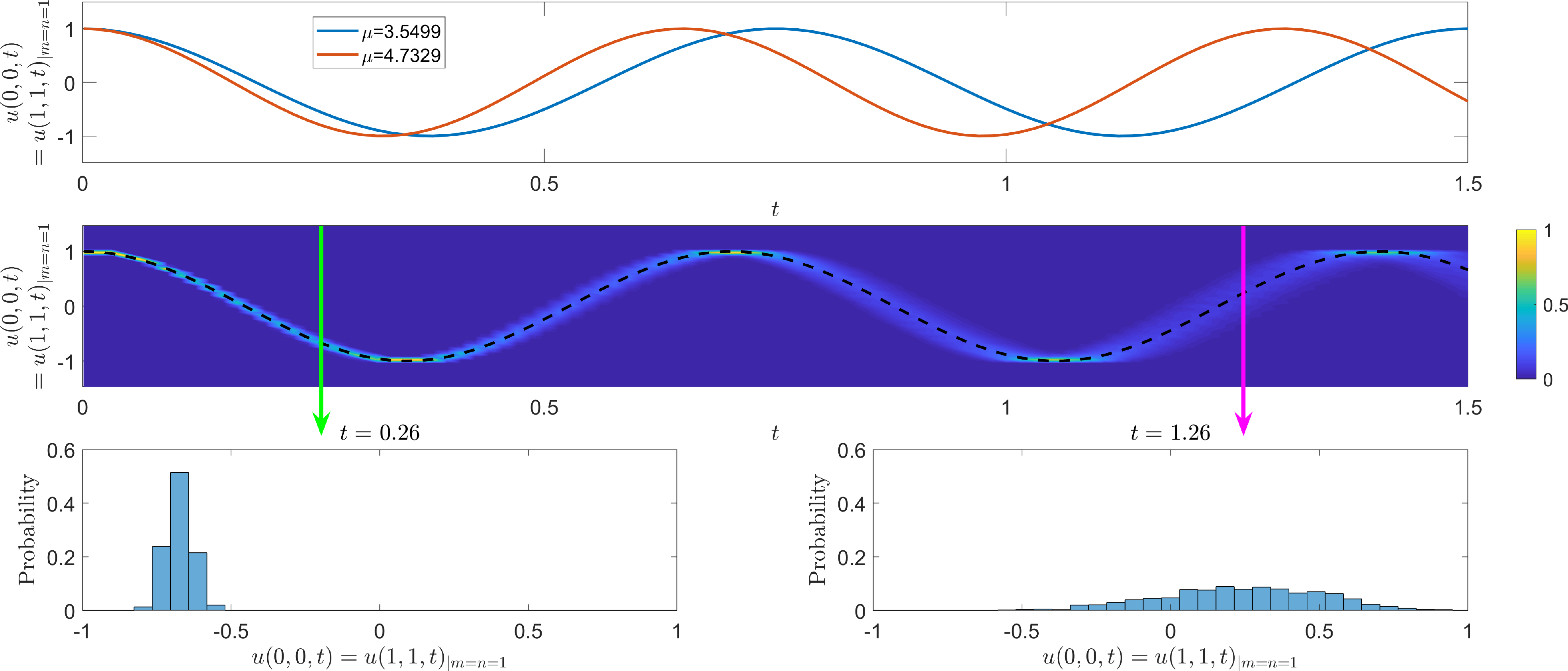}
		\caption{Stochastic displacement $u(X,Y,t)$ of the edges $(X,Y,Z)\in\{(0,0,Z), (1,1,Z)\}$ of the cuboid in dynamic generalised shear, when $m=n=1$, $A_{11}=1$, $B_{11}=0$, $\rho=1$, and $\mu$ is drawn from the Gamma distribution with $\rho_{1}=405$ and $\rho_{2}=0.01$. The top figure illustrates the displacement over time of two cuboids, with randomly chosen values of $\mu$, derived from the specified Gamma distribution. The middle figure illustrates a probability histogram at each time instant. Specifically, the integral of the probabilities over the displacements at any given time instant is equal to $1$. The histogram comprises of $1000$ stochastic simulations and the colour bar defines the probability of finding a given displacement at a given time. The dashed black line corresponds to the expected values based only on mean value, $\underline{\mu}=\rho_{1}\rho_{2}=4.05$. The bottom two figures illustrate specific histogram distributions at two given times (noted above each figure). These are the distributions that would be seen if the middle figure was cut along the green and magenta arrows, respectively.}\label{fig:stochwave-cube}
	\end{center}
\end{figure}

In both the above cases, (i) and (ii), respectively, by \eqref{eq:cube:1st},
\begin{equation}\label{eq:cube:motion}
\ddot{u}=\frac{\mu}{\rho}\left(u_{XX}+u_{YY}\right).
\end{equation}
It remains to solve, by standard procedures, the linear wave equation \eqref{eq:cube:motion}, describing the propagation of waves, subject to the given initial and boundary conditions. To solve this equation, we let the shear stresses $T_{xz}$ and $T_{yz}$, defined by \eqref{eq:cube:stresses}, vanish at the sides, i.e.,
\begin{equation}\label{eq:cube:bcs}
\begin{split}
T_{xz}(0,Y,Z,t)=T_{xz}(1,Y,Z,t)=0\qquad&\iff\qquad u_{X}(0,Y,t)=u_{X}(1,Y,t)=0,\\
T_{yz}(X,0,Z,t)=T_{yz}(X,1,Z,t)=0\qquad&\iff\qquad u_{Y}(X,0,t)=u_{Y}(X,1,t)=0.
\end{split}
\end{equation}
In this case, the general solution takes the form
\begin{equation}\label{eq:cube:sol}
u(X,Y,t)=\sum_{m=1}^{\infty}\sum_{n=1}^{\infty}\left[A_{mn}\cos\left(\omega_{mn}t\right)+B_{mn}\sin\left(\omega_{mn}t\right)\right]\cos\left(\pi mX\right)\cos\left(\pi nY\right),
\end{equation}
where
\begin{equation}\label{eq:cube:omegam}
\omega_{mn}=\pi\sqrt{\left(m^2+n^2\right)\frac{\mu}{\rho}},
\end{equation}
and
\begin{eqnarray}
&&A_{mn}=4\int_{0}^{1}\left[\int_{0}^{1}u_{0}(X,Y)\cos\left(\pi mX\right)dX\right]\cos\left(\pi nY\right)dY,\\
&&B_{mn}=\frac{4}{\omega_{mn}}\int_{0}^{1}\left[\int_{0}^{1}\dot{u}_{0}(X,Y)\cos\left(\pi mX\right)dX\right]\cos\left(\pi nY\right)dY.
\end{eqnarray}
These oscillations under the generalised shear motion \eqref{eq:cube:gshear} cannot be completely `free', due to the non-zero tractions corresponding to the cases (i) and (ii), respectively. Note that the condition \eqref{eq:cube:curl} is not satisfied.

As $\mu$ is a random variable, it follows that the speed of wave propagation, $\sqrt{\mu/\rho}$, is stochastic. Hence, both the period and the amplitude of the oscillations are stochastic. As an example, we consider the initial data $u_{0}(X,Y)=\cos(\pi X)\cos(\pi Y)$ and $\dot{u}_{0}(X,Y)=0$ leading to $A_{11}=1$ and $B_{11}=0$. In Figure~\ref{fig:stochwave-cube}, we illustrate the stochastic dynamic displacement on the edges $(X,Y,Z)\in\{(0,0,Z), (1,1,,Z)\}$ when $m=n=1$, $A_{11}=1$, $B_{11}=0$, $\rho=1$, and $\mu$ is drawn from the Gamma distribution with hyperparameters $\rho_{1}=405$ and $\rho_{2}=0.01$, as represented in Figure~\ref{fig:mu-gpdf}. The top plot of Figure~\ref{fig:stochwave-cube} represents two single simulations, with two different values of $\mu$ drawn from the distribution, illustrating the variety of outcomes that can be obtained. The middle plot of Figure~\ref{fig:stochwave-cube} then represents histograms of the ensemble data. Namely, since not all material parameters are equally likely, not all outcomes are equally likely. Specifically, the values of $u(0,0,t)$ are most likely going to be near the mean value (dashed line) with the probability of observing alternative values of $u$ decreasing as we tend away from the mean. We note from Figure~\ref{fig:stochwave-cube} that, as we might expect, extremal probabilities always occur at the extreme displacement of the oscillations, i.e., when the cuboid is slowest. However, in between these probability maxima, the variance grows over time. Thus, although the displacements are initially close (seen explicitly in the top of Figure~\ref{fig:stochwave-cube} and by the tight distribution around the mean in the bottom left of Figure~\ref{fig:stochwave-cube}), eventually, the phase difference dominates causing the displacements to diverge (top of Figure~\ref{fig:stochwave-cube}), and an increase in the variance of the distribution (bottom right of Figure~\ref{fig:stochwave-cube}).

\section{Quasi-equilibrated radial-axial motion of a stochastic hyperelastic cylindrical tube}\label{sec:tube}

In this section, we analyse the stability and finite amplitude oscillations of a stochastic hyperelastic cylindrical tube subject to the combined radial and axial quasi-equilibrated dynamic deformation.

\subsection{Dynamic radial-axial deformation of a cylindrical tube}

\begin{figure}[htbp]
	\begin{center}
		\includegraphics[width=0.65\textwidth]{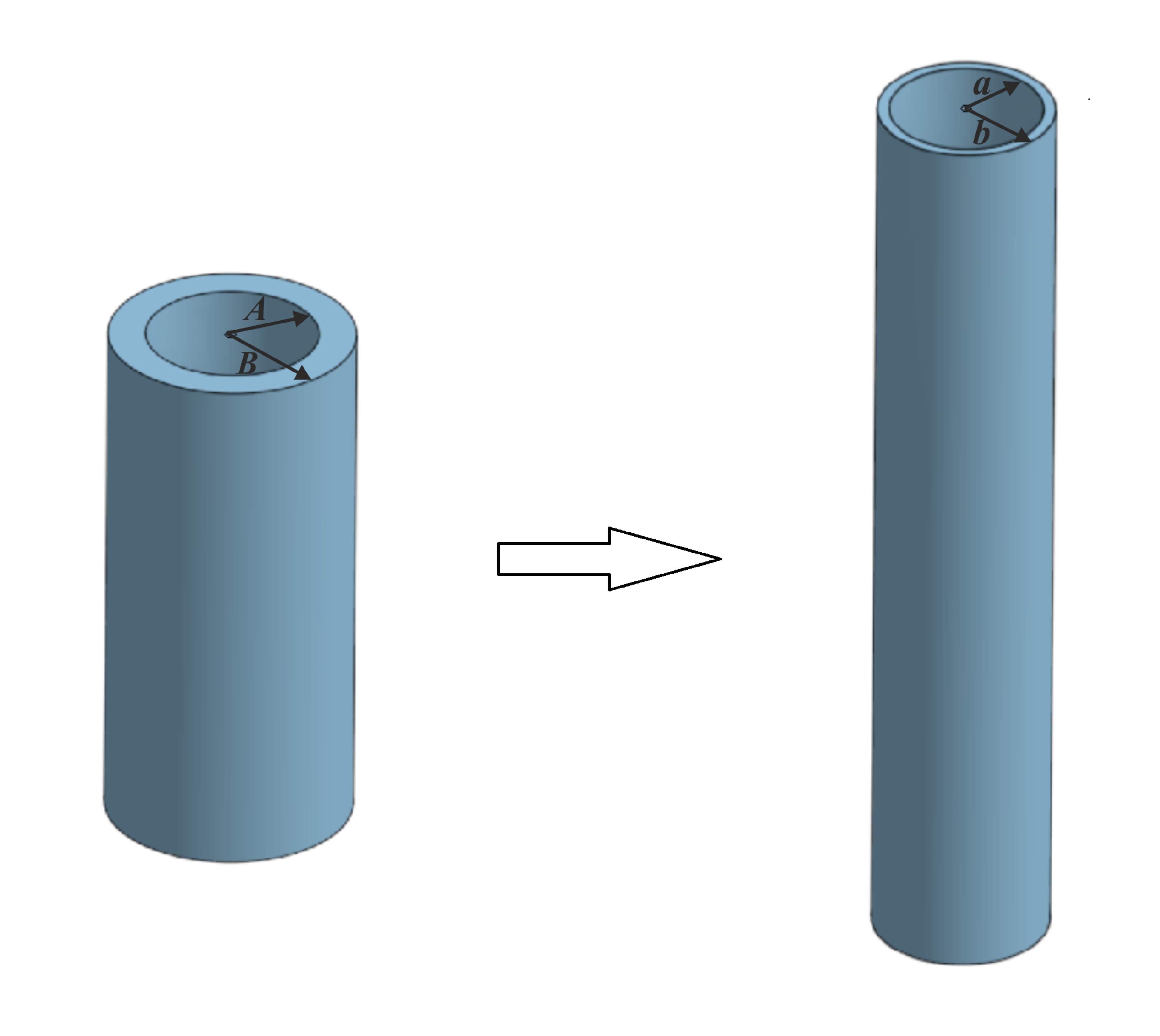}
		\caption{Schematic of inflation of a cylindrical tube, showing the reference state, with inner radius $A$ and outer radius $B$ (left), and the deformed state, with inner radius $a$ and outer radius $b$ (right), respectively.}\label{fig:tube}
	\end{center}
\end{figure}

For a circular cylindrical tube, the combined radial and axial motion is described by (see Figure~\ref{fig:tube})
\begin{equation}\label{eq:tube:deform}
r^2=a^2+\frac{R^2-A^2}{\alpha},\qquad \theta=\Theta,\qquad z=\alpha Z,
\end{equation}
where $(R,\Theta,Z)$ and $(r,\theta,z)$ are the cylindrical polar coordinates in the reference and current configuration, respectively, such that $A\leq R\leq B$, $A$ and $B$ are the inner and outer radii in the undeformed state, respectively, $a=a(t)$  and $b=b(t)=\sqrt{a^2+\left(B^2-A^2\right)/\alpha}$ are the inner and outer radius at time $t$, respectively, and $\alpha>0$ is a given constant (when $\alpha<0$, the tube is everted, so that the inner surface becomes the outer surface). When $\alpha=1$, the time-dependent deformation \eqref{eq:tube:deform} simplifies to that studied also in \cite{Beatty:2007,Knowles:1960,Knowles:1962}. The case when $\alpha$ is time-dependent was considered in \cite{Shahinpoor:1973}.

The radial-axial motion \eqref{eq:tube:deform} of the cylindrical tube is fully determined by the inner radius $a$ at time $t$, which in turn is obtained from the initial conditions. Thus, the acceleration $\ddot{r}$ can be computed in terms of the acceleration $\ddot{a}$ on the inner surface. By the governing equations \eqref{eq:tube:deform}, the condition \eqref{eq:ddotx:curl} is valid for $\textbf{x}=(r,\theta,z)^{T}$, since
\begin{equation}\label{eq:tube:curl}
\textbf{0}=\mathrm{curl}\ \ddot{\textbf{x}}=
\left[
\begin{array}{c}
(\partial\ddot{z}/\partial\theta)/r-\partial\ddot{\theta}/\partial z\\
\partial\ddot{r}/\partial z-\partial\ddot{z}/\partial r\\
\partial\ddot{\theta}/\partial r-(\partial\ddot{r}/\partial\theta)/r
\end{array}
\right],
\end{equation}
and the acceleration potential, $\xi$, satisfies \eqref{eq:cp}. Hence, this is a quasi-equilibrated motion, such that
\begin{equation}\label{eq:tube:cp}
-\frac{\partial\xi}{\partial r}=\ddot{r}=\frac{\dot{a}^2}{r}+\frac{a\ddot{a}}{r}-\frac{a^2\dot{a}^2}{r^3},
\end{equation}
and, by integrating \eqref{eq:tube:cp}, the acceleration potential, $\xi$, is given by \cite[p.~215]{TruesdellNoll:2004}
\begin{equation}\label{eq:tube:xi}
-\xi=\dot{a}^2\log r+a\ddot{a}\log r+\frac{a^2\dot{a}^2}{2r^2}=\dot{r}^2\log r+r\ddot{r}\log r+\frac{1}{2}\dot{r}^2.
\end{equation}
The deformation gradient of \eqref{eq:tube:deform}, with respect to the polar coordinates $(R,\Theta,Z)$, is equal to
\begin{equation}\label{eq:tube:F}
\textbf{F}=\mathrm{diag}\left(\frac{R}{\alpha r}, \frac{r}{R}, \alpha\right),
\end{equation}
the Cauchy-Green deformation tensor is
\begin{equation}\label{eq:tube:B}
\textbf{B}=\textbf{F}^2=\mathrm{diag}\left(\frac{R^2}{\alpha^2 r^2}, \frac{r^2}{R^2}, \alpha^2\right),
\end{equation}
and the principal invariants take the form
\begin{equation}\label{eq:tube:I123}
\begin{split}
I_{1}=&\mathrm{tr}\ (\textbf{B})=\frac{R^2}{\alpha^2 r^2}+\frac{r^2}{R^2}+\alpha^2,\\
I_{2}=&\frac{1}{2}\left[\left(\mathrm{tr}\,\textbf{B}\right)^{2}-\mathrm{tr}\left(\textbf{B}^{2}\right)\right]=\frac{\alpha^2r^2}{R^2}+\frac{R^2}{r^2}+\frac{1}{\alpha^2},\\
I_{3}=&\det\textbf{B}=1.
\end{split}
\end{equation}
Thus, the principal components of the equilibrium Cauchy stress tensor at time $t$ are
\begin{equation}\label{eq:tube:stresses}
\begin{split}
T^{(0)}_{rr}&=-p^{(0)}+\beta_{1}\frac{R^2}{\alpha^2r^2}+\beta_{-1}\frac{\alpha^2r^2}{R^2},\\
T^{(0)}_{\theta\theta}&=T^{(0)}_{rr}+\left(\beta_{1}-\beta_{-1}\alpha^2\right)\left(\frac{r^2}{R^2}-\frac{R^2}{\alpha^2r^2}\right),\\
T^{(0)}_{zz}&=T^{(0)}_{rr}+\left(\beta_{1}-\beta_{-1}\frac{r^2}{R^2}\right)\left(\alpha^2-\frac{R^2}{\alpha^2r^2}\right),
\end{split}
\end{equation}
where $p^{(0)}$ is the Lagrangian multiplier for the incompressibility constraint ($I_{3}=1$), and
\begin{equation}\label{eq:tube:betas}
\beta_{1}=2\frac{\partial W}{\partial I_{1}},\qquad \beta_{-1}=-2\frac{\partial W}{\partial I_{2}}
\end{equation}
are the nonlinear material parameters, with $I_{1}$  and $I_{2}$ given by \eqref{eq:tube:I123}.

As the stress components depend only on the radius $r$, the system of equilibrium equations reduces to
\begin{equation}\label{eq:tube:equilibrium}
\frac{\partial T^{(0)}_{rr}}{\partial r}=\frac{T^{(0)}_{\theta\theta}-T^{(0)}_{rr}}{r}.
\end{equation}
Hence, by \eqref{eq:tube:stresses} and \eqref{eq:tube:equilibrium}, the radial Cauchy stress for the equilibrium state at time $t$ is equal to
\begin{equation}\label{eq:tube:T0rr}
T^{(0)}_{rr}(r,t)=\int\left(\beta_{1}-\beta_{-1}\alpha^2\right)\left(\frac{r^2}{R^2}-\frac{R^2}{\alpha^2r^2}\right)\frac{dr}{r}+\psi(t),
\end{equation}
where $\psi=\psi(t)$ is an arbitrary function of time. Substitution of \eqref{eq:tube:xi} and \eqref{eq:tube:T0rr} into \eqref{eq:T} then gives the principal Cauchy stress components at time $t$ as follows,
\begin{equation}\label{eq:tube:Trr}
\begin{split}
T_{rr}(r,t)&=\rho\left(a\ddot{a}\log r+\dot{a}^2\log r+\frac{a^2\dot{a}^2}{2r^2}\right)+\int\left(\beta_{1}-\beta_{-1}\alpha^2\right)\left(\frac{r^2}{R^2}-\frac{R^2}{\alpha^2r^2}\right)\frac{dr}{r}+\psi(t),\\
T_{\theta\theta}(r,t)&=T_{rr}(r,t)+\left(\beta_{1}-\beta_{-1}\alpha^2\right)\left(\frac{r^2}{R^2}-\frac{R^2}{\alpha^2r^2}\right),\\
T_{zz}(r,t)&=T_{rr}(r,t)+\left(\beta_{1}-\beta_{-1}\frac{r^2}{R^2}\right)\left(\alpha^2-\frac{R^2}{\alpha^2r^2}\right).
\end{split}
\end{equation}
In \eqref{eq:tube:Trr}, the function $\beta_{1}-\alpha^2\beta_{-1}$ can be interpreted as the following nonlinear shear modulus \cite{Mihai:2017:MG}
\begin{equation}\label{eq:tube:shearmod}
\widetilde{\mu}=\beta_{1}-\beta_{-1}\alpha^2,
\end{equation}
corresponding to the combined deformation of simple shear superposed on axial stretch, described by \eqref{eq:stretchshear}, with shear parameter $k=\sqrt{\alpha^2R^2/r^2+\alpha^4r^2/R^2-\alpha^6-1}$ and stretch parameter $\alpha$. As shown in \cite{Mihai:2017:MG}, this modulus is positive if the BE inequalities \eqref{eq:BE} hold. In this case, the integrand is negative for $0<r^2/R^2<1/\alpha$ and positive for $r^2/R^2>1/\alpha$. Using the first equation in \eqref{eq:tube:deform}, it is straightforward to show that $0<r^2/R^2<1/\alpha$ (respectively, $r^2/R^2>1/\alpha$) is equivalent to $0<a^2/A^2<1/\alpha$ (respectively, $a^2/A^2>1/\alpha$). When $\alpha=1$, the modulus defined by \eqref{eq:tube:shearmod} coincides with the generalised shear modulus defined in \cite[p.~174]{TruesdellNoll:2004}, and also in \cite{Beatty:2007}.

In the limiting case when $\alpha\to 1$ and $k\to 0$, the nonlinear shear modulus defined by \eqref{eq:tube:shearmod} converges to the classical shear modulus from the infinitesimal theory \cite{Mihai:2017:MG},
\begin{equation}\label{eq:tube:shearmod:lin}
\mu=\lim_{\alpha\to1}\lim_{k\to0}\widetilde{\mu}.
\end{equation}
In this case, as $R^2/r^2\to1$, the three stress components defined by \eqref{eq:tube:Trr} are equal.

Next, for the cylindrical tube deforming by \eqref{eq:tube:deform}, we set the inner and outer radial pressures acting on the curvilinear surfaces $r=a(t)$ and $r=b(t)$ at time $t$ (measured per unit area in the present configuration), as $T_{1}(t)$ and $T_{2}(t)$, respectively \cite[pp.~214-217]{TruesdellNoll:2004}. Evaluating $T_{1}(t)=-T_{rr}(a,t)$ and $T_{2}(t)=-T_{rr}(b,t)$, using \eqref{eq:tube:Trr}, with $r=a$ and $r=b$, respectively, then subtracting the results, then gives
\begin{equation}\label{eq:tube:P1P2:r}
\begin{split}
T_{1}(t)-T_{2}(t)&=\frac{\rho}{2}\left[\left(a\ddot{a}+\dot{a}^2\right)\log\frac{b^2}{a^2}+\dot{a}^2\left(\frac{a^2}{b^2}-1\right)\right]
+\int_{a}^{b}\widetilde{\mu}\left(\frac{r^2}{R^2}-\frac{R^2}{\alpha^2r^2}\right)\frac{dr}{r}\\
&=\frac{\rho A^2}{2}\left[\left(\frac{a}{A}\frac{\ddot{a}}A+\frac{\dot{a}^2}{A^2}\right)\log\frac{b^2}{a^2}+\frac{\dot{a}^2}{A^2}\left(\frac{a^2}{b^2}-1\right)\right]
+\int_{a}^{b}\widetilde{\mu}\left(\frac{r^2}{R^2}-\frac{R^2}{\alpha^2r^2}\right)\frac{dr}{r}.
\end{split}
\end{equation}
Setting the notation
\begin{equation}\label{eq:tube:uxgamma}
u=\frac{r^2}{R^2}=\frac{r^2}{\alpha\left(r^2-a^2\right)+A^2},\qquad
x=\frac{a}{A},\qquad
\gamma=\frac{B^2}{A^2}-1,
\end{equation}
we can rewrite
\[
\begin{split}
\left(\frac{a}{A}\frac{\ddot{a}}A+\frac{\dot{a}^2}{A^2}\right)\log\frac{b^2}{a^2}+\frac{\dot{a}^2}{A^2}\left(\frac{a^2}{b^2}-1\right)
&=\left(\ddot{x}x+\dot{x}^2\right)\log\left(1+\frac{\gamma}{\alpha x^2}\right)-\dot{x}^2\frac{\frac{\gamma}{\alpha x^2}}{1+\frac{\gamma}{\alpha x^2}}\\
&=\frac{1}{2x}\frac{d}{dx}\left[\dot{x}^2x^2\log\left(1+\frac{\gamma}{\alpha x^2}\right)\right]
\end{split}
\]
and
\[
\begin{split}
\int_{a}^{b}\widetilde{\mu}\left(\frac{r^2}{R^2}-\frac{R^2}{\alpha^2r^2}\right)\frac{dr}{r}
&=\int_{a}^{b}\widetilde{\mu}\left[\frac{r^2}{\alpha\left(r^2-a^2\right)+A^2}-\frac{\alpha\left(r^2-a^2\right)+A^2}{\alpha^2r^2}\right]\frac{dr}{r}\\
&=\frac{1}{2}\int_{\frac{x^2+\frac{\gamma}{\alpha}}{1+\gamma}}^{x^2}\widetilde{\mu}\frac{1+\alpha u}{\alpha^2u^2}du.
\end{split}
\]
Then, we can express the equation \eqref{eq:tube:P1P2:r} equivalently as follows,
\begin{equation}\label{eq:tube:P1P2:x}
2x\frac{T_{1}(t)-T_{2}(t)}{\rho A^2}=\frac{1}{2}\frac{d}{dx}\left[\dot{x}^2x^2\log\left(1+\frac{\gamma}{\alpha x^2}\right)\right]
+\frac{x}{\rho A^2}\int_{\frac{x^2+\frac{\gamma}{\alpha}}{1+\gamma}}^{x^2}\widetilde{\mu}\frac{1+\alpha u}{\alpha^2u^2}du.
\end{equation}
Note that, when the BE inequalities  \eqref{eq:BE} hold, $\widetilde{\mu}>0$, and the integral in \eqref{eq:tube:P1P2:r}, or equivalently in \eqref{eq:tube:P1P2:x}, is negative if $0<u<1/\alpha$ (i.e., if $0<x<1/\sqrt{\alpha}$) and positive if $u>1/\alpha$ (i.e., if $x>1/\sqrt{\alpha}$).

In the static case, where $\dot{a}=0$ and $\ddot{a}=0$, \eqref{eq:tube:P1P2:r} becomes
\begin{equation}\label{eq:tube:P1P2:r:static}
T_{1}(t)-T_{2}(t)=\int_{a}^{b}\widetilde{\mu}\left(\frac{r^2}{R^2}-\frac{R^2}{\alpha^2r^2}\right)\frac{dr}{r},
\end{equation}
and \eqref{eq:tube:P1P2:x} reduces to
\begin{equation}\label{eq:tube:P1P2:x:static}
	2\frac{T_{1}(t)-T_{2}(t)}{\rho A^2}=\frac{1}{\rho A^2}\int_{\frac{x^2+\frac{\gamma}{\alpha}}{1+\gamma}}^{x^2}\widetilde{\mu}\frac{1+\alpha u}{\alpha^2u^2}du.
\end{equation}

For the cylindrical tube in finite dynamic deformation, we set
\begin{equation}\label{eq:tube:G}
G(x,\gamma)=\frac{1}{\rho A^2}\int_{1/\sqrt{\alpha}}^{x}\left(\zeta\int_{\frac{\zeta^2+\frac{\gamma}{\alpha}}{1+\gamma}}^{\zeta^2}\widetilde{\mu}\frac{1+\alpha u}{\alpha^2u^2}du\right)d\zeta,
\end{equation}
and find that $G(x,\gamma)$ is monotonically decreasing when $0<x<1/\sqrt{\alpha}$ and increasing when $x>1/\sqrt{\alpha}$. This function will be useful in establishing whether the radial motion is oscillatory or not.

We also set the pressure impulse (suddenly applied pressure difference)
\begin{equation}\label{eq:tube:impulse}
2\alpha\frac{T_{1}(t)-T_{2}(t)}{\rho A^2}=\left\{
\begin{array}{cc}
0 & \mbox{if}\ t\leq0,\\
p_{0} & \mbox{if}\ t>0,
\end{array}
\right.
\end{equation}
where $p_{0}$ is constant in time. Then, integrating \eqref{eq:tube:P1P2:x} once gives
\begin{equation}\label{eq:tube:ode}
\frac{1}{2}\dot{x}^2x^2\log\left(1+\frac{\gamma}{\alpha x^2}\right)+G(x,\gamma)=\frac{p_{0}}{2\alpha}\left(x^2-\frac{1}{\alpha}\right)+C,
\end{equation}
with $G(x,\gamma)$ defined by \eqref{eq:tube:G} and
\begin{equation}\label{eq:tube:C}
C=\frac{1}{2}\dot{x}_{0}^2x_{0}^2\log\left(1+\frac{\gamma}{\alpha x_{0}^2}\right)+G(x_{0},\gamma)-\frac{p_{0}}{2\alpha}\left(x_{0}^2-\frac{1}{\alpha}\right),
\end{equation}
where $x(0)=x_{0}$ and $\dot{x}(0)=\dot{x}_{0}$ are the initial conditions. By \eqref{eq:tube:ode},
\begin{equation}\label{eq:tube:dotx}
\dot{x}=\pm\sqrt{\frac{\frac{p_{0}}{\alpha}\left(x^2-\frac{1}{\alpha}\right)+2C-2G(x,\gamma)}{x^2\log\left(1+\frac{\gamma}{\alpha x^2}\right)}}.
\end{equation}
Physically, this system is analogous to the motion of a point mass with energy
\begin{equation}
E=\frac{1}{2} m(x)\dot x^{2}+V(x).
\end{equation}
The energy is $E=C$, the potential is given by $V(x)=G(x,\gamma)-\frac{p_{0}}{2\alpha}\left(x^2-\frac{1}{\alpha}\right)$ and the position-dependent mass is $m(x)={x^2\log\left(1+\frac{\gamma}{\alpha x^2}\right)}$. Due to the constraints on the function $G$, this system has  simple dynamics. Depending on the constant $\mu$, the system may have a static state or periodic motion. 
Indeed, the radial motion is periodic if and only if the following equation,
\begin{equation}\label{eq:tube:GC}
G(x,\gamma)=\frac{p_{0}}{2\alpha}\left(x^2-\frac{1}{\alpha}\right)+C,
\end{equation}
has exactly two distinct solutions, representing the amplitudes of the oscillation, $x=x_{1}$ and $x=x_{2}$, such that $0<x_{1}<x_{2}<\infty$. Then, by \eqref{eq:tube:uxgamma}, the minimum and maximum radii of the inner surface in the oscillation are equal to $x_{1}A$ and $x_{2}A$, respectively, and by \eqref{eq:tube:dotx}, the period of oscillation is equal to
\begin{equation}\label{eq:tube:T}
T=2\left|\int_{x_{1}}^{x_{2}}\frac{dx}{\dot{x}}\right|=2\left|\int_{x_{1}}^{x_{2}}\sqrt{\frac{x^2\log\left(1+\frac{\gamma}{\alpha x^2}\right)}{\frac{p_{0}}{\alpha}\left(x^2-\frac{1}{\alpha}\right)+2C-2G(x,\gamma)}}dx\right|.
\end{equation}
Note that both the amplitudes and period of the oscillation are random variables described in terms of probability distributions.

\subsection{Radial oscillations of a cylindrical tube of stochastic Mooney-Rivlin material}

For cylindrical tubes of stochastic Mooney-Rivlin material defined by \eqref{eq:W:stoch}, with $\mu=\mu_{1}+\mu_{2}>0$, evaluating the integral in \eqref{eq:tube:G} gives (see Appendix~\ref{sec:append} for detailed calculations)
\begin{equation}\label{eq:tube:G:MR}
G(x,\gamma)=\frac{\widetilde{\mu}}{2\alpha\rho A^2}\left(x^2-\frac{1}{\alpha}\right)\log\frac{1+\gamma}{1+\frac{\gamma}{\alpha x^2}},
\end{equation}
where $\widetilde{\mu}=\mu_{1}+\mu_{2}\alpha^2$. In this case,  assuming that the nonlinear shear modulus $\mu$ has a uniform lower bound, i.e.,
\begin{equation}\label{eq:tube:shearnod:bound}
\mu>\eta,
\end{equation}
for some constant $\eta>0$, it follows that
\begin{equation}\label{eq:tube:G:infty}
\lim_{x\to0}G(x,\gamma)= \lim_{x\to\infty}G(x,\gamma)=\infty.
\end{equation}

\paragraph{(i)} If $p_{0}=0$ and $C>0$, then equation \eqref{eq:tube:GC} has exactly two solutions, $x=x_{1}$ and $x=x_{2}$, satisfying $0<x_{1}<1/\sqrt{\alpha}<x_{2}<\infty$, for any positive constant $C$. It should be noted that, by \eqref{eq:tube:Trr}, if $T_{rr}(r,t)=0$ at $r=a$ and $r=b$, so that $T_{1}(t)=T_{2}(t)=0$, then
$T_{\theta\theta}(r,t)\neq 0$ and $T_{zz}(r,t)\neq 0$ at $r=a$ and $r=b$, unless $\alpha\to1$ and $r^2/R^2\to1$. Thus, in general, these oscillations cannot be `free' \cite{Shahinpoor:1973}.

\begin{figure}[htbp]
	\begin{center}
		\includegraphics[width=0.45\textwidth]{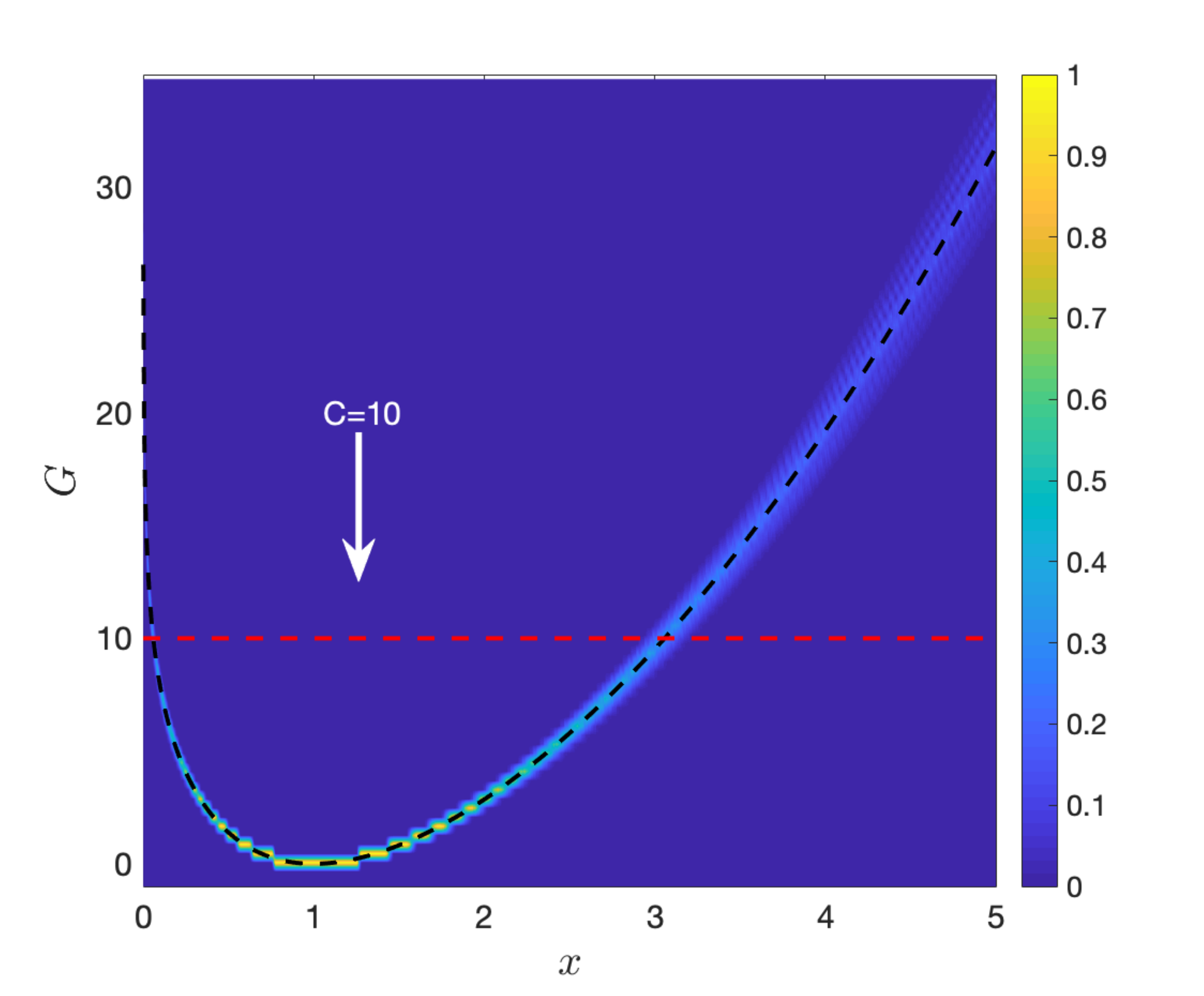}\qquad
		\includegraphics[width=0.45\textwidth]{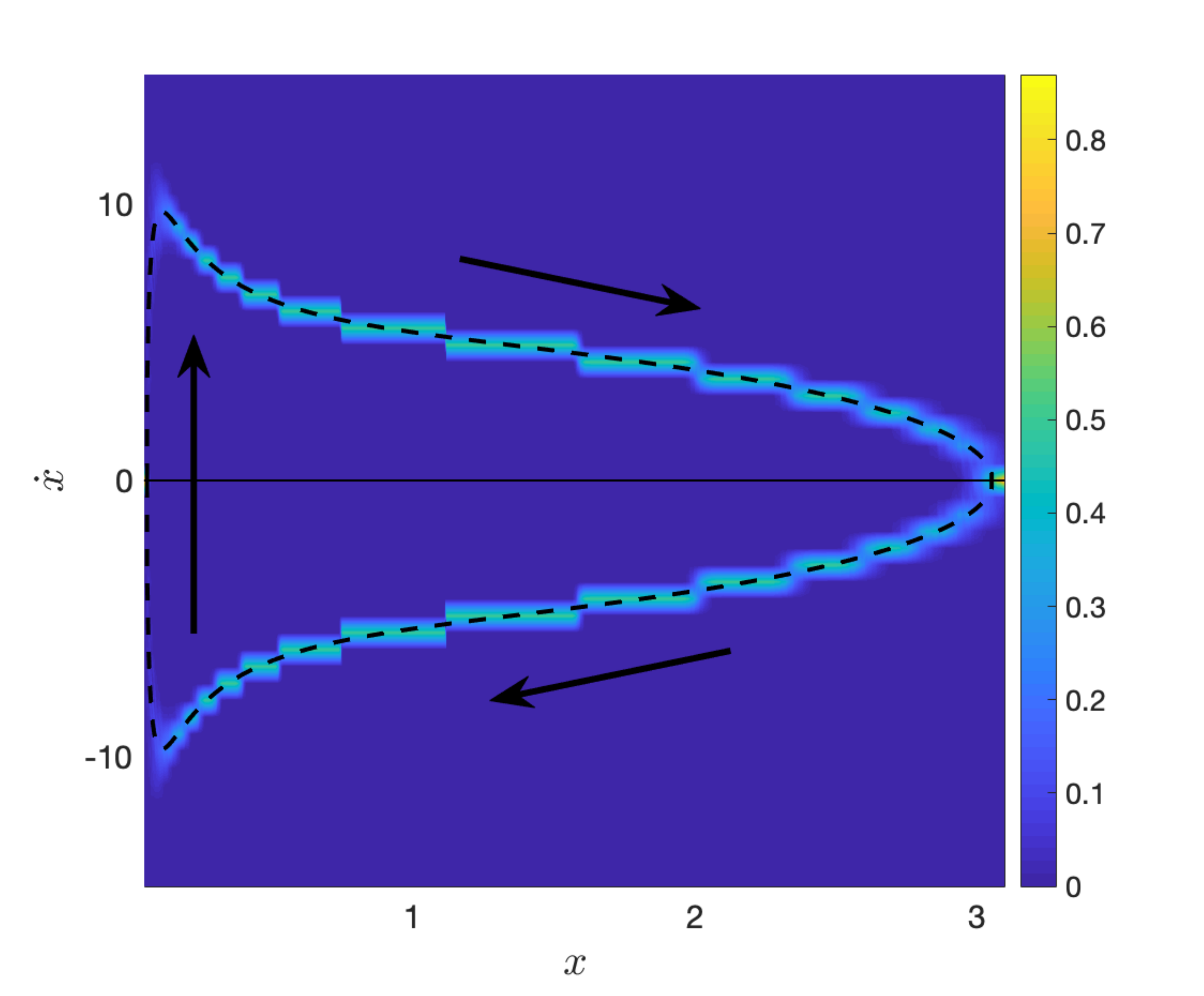}
		\caption{The function $G(x,\gamma)$, defined by \eqref{eq:tube:G:MR}, intersecting the (dashed red) line $C=10$ when $p_{0}=0$ (left), and the associated velocity, given by \eqref{eq:tube:dotx} (right), for a cylindrical tube of stochastic Mooney-Rivlin material when $\alpha=1$, $\rho=1$, $A=1$, $\gamma=1$, and $\widetilde{\mu}=\mu=\mu_1+\mu_{2}$ is drawn from the Gamma distribution with $\rho_{1}=405$ and $\rho_{2}=0.01$. The dashed black lines correspond to the expected values based only on mean value, $\underline{\mu}=\rho_{1}\rho_{2}=4.05$. Each distribution was calculated from the average of $1000$ stochastic simulations.}\label{fig:stoch-freetube}
	\end{center}
\end{figure}

In Figure~\ref{fig:stoch-freetube}, for example, we represent the stochastic function $G(x,\gamma)$, defined by \eqref{eq:tube:G:MR}, intersecting the line $C=10$ , to solve equation \eqref{eq:tube:GC} when $p_{0}=0$, and the associated velocity, given by \eqref{eq:tube:dotx},  assuming that $\alpha=1$, $\rho=1$, $A=1$, $\gamma=1$, and $\mu$ follows the Gamma distribution with hyperparameters $\rho_{1}=405$ and $\rho_{2}=0.01$ (see Figure~\ref{fig:mu-gpdf}).

- For a thin-walled tube \cite{Knowles:1960,Shahinpoor:1971:SN}, where $\alpha=1$ and $\gamma\to 0$, equation \eqref{eq:tube:ode} takes the form
\begin{equation}\label{eq:tube:ode:MR:thin}
\dot{x}^2+\frac{\mu}{\rho A^2}\left(x^2+\frac{1}{x^2}\right)=\dot{x}_{0}^2+\frac{\mu}{\rho A^2}\left(x_{0}^2+\frac{1}{x_{0}^2}\right),
\end{equation}
and has the explicit solution \cite{Shahinpoor:1971:SN}
\begin{equation}\label{eq:tube:x:MR:thin}
x=\sqrt{\left[x_{0}\cos\left(\frac{t}{A}\sqrt{\frac{\mu}{\rho}}\right)+\dot{x}_{0}A\sqrt{\frac{\rho}{\mu}}\sin\left(\frac{t}{A}\sqrt{\frac{\mu}{\rho}}\right)\right]^2+\frac{1}{x_{0}^2}\sin^2\left(\frac{t}{A}\sqrt{\frac{\mu}{\rho}}\right)}.
\end{equation}
In this case, assuming that the shear modulus, $\mu$, has a uniform lower bound, equation \eqref{eq:tube:GC}  becomes \cite{Knowles:1960}
\begin{equation}\label{eq:tube:GC:MR:thin}
x^2+\frac{1}{x^2}=\frac{\rho A^2}{\mu}\dot{x}_{0}^2+x_{0}^2+\frac{1}{x_{0}^2}.
\end{equation}
This equation can be solved directly to find the amplitudes
\begin{equation}\label{eq:tube:amp:MR:thin}
x_{1,2}=\sqrt{\frac{\frac{\rho A^2}{\mu}\dot{x}_{0}^2+x_{0}^2+\frac{1}{x_{0}^2}\pm\sqrt{\left(\frac{\rho A^2}{\mu}\dot{x}_{0}^2+x_{0}^2+\frac{1}{x_{0}^2}\right)^2-4}}{2}}.
\end{equation}
Noting that $x_{2}=1/x_{1}$, the period of the oscillations can be calculated as
\begin{equation}\label{eq:tube:T:MR:thin}
T=2\sqrt{\frac{\rho A^2}{\mu}}\left|\int_{x_{1}}^{1/x_{1}}\frac{dx}{\sqrt{\frac{\rho A^2}{\mu}\dot{x}_{0}^2+x_{0}^2+\frac{1}{x_{0}^2}-x^2-\frac{1}{x^2}}}\right|=\pi A\sqrt{\frac{\rho}{\mu}}.
\end{equation}

\begin{figure}[htbp]
	\begin{center}
		\includegraphics[width=\textwidth]{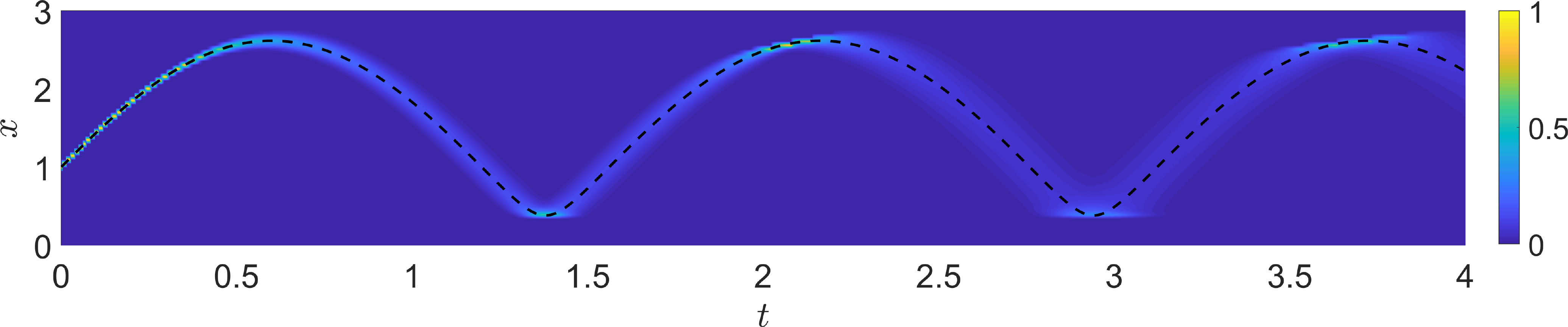}
		\caption{Stochastic solution given by \eqref{eq:tube:x:MR:thin}, with the initial conditions $x_{0}=1$ and $\dot{x}_{0}=4.5$, for a thin-walled tube, where $\rho=1$, $A=1$, and $\mu$ is drawn from the Gamma distribution with $\rho_{1}=405$ and $\rho_{2}=0.01$. The dashed black line corresponds to the expected values based only on mean value, $\underline{\mu}=\rho_{1}\rho_{2}=4.05$. The distribution was calculated from the average of $1000$ stochastic simulations.}\label{fig:stochwave-freethintube}
	\end{center}
\end{figure}

In Figure~\ref{fig:stochwave-freethintube}, we illustrate the stochastic solution given by \eqref{eq:tube:x:MR:thin}, with the initial conditions $x_{0}=1$ and $\dot{x}_{0}=4.5$, assuming that $\rho=1$, $A=1$, and $\mu$ satisfies the Gamma distribution with hyperparameters $\rho_{1}=405$ and $\rho_{2}=0.01$.

\paragraph{(ii)} When $p_{0}\neq 0$ and $C\geq 0$, substitution of \eqref{eq:tube:G:MR} in \eqref{eq:tube:GC} gives
\begin{equation}\label{eq:tube:p0:MR}
p_{0}=\frac{\widetilde{\mu}}{\rho A^2}\log\frac{1+\gamma}{1+\frac{\gamma}{\alpha x^2}}-\frac{2\alpha C}{x^2-\frac{1}{\alpha}}.
\end{equation}
as the right-hand side of the above equation is a monotonically increasing function of $x$, there exists a unique positive $x$ satisfying \eqref{eq:tube:p0:MR} if and only if the following condition holds,
\begin{equation*}\label{eq:tube:p0:MR:limbounds}
\lim_{x\to0}\left(\frac{\widetilde{\mu}}{\rho A^2}\log\frac{1+\gamma}{1+\frac{\gamma}{\alpha x^2}}-\frac{2\alpha C}{x^2-\frac{1}{\alpha}}\right)<p_{0}<\lim_{x\to\infty}\left(\frac{\widetilde{\mu}}{\rho A^2}\log\frac{1+\gamma}{1+\frac{\gamma}{\alpha x^2}}-\frac{2\alpha C}{x^2-\frac{1}{\alpha}}\right),
\end{equation*}
that is,
\begin{equation}\label{eq:tube:p0:MR:bounds}
-\infty<p_{0}<\frac{\widetilde{\mu}}{\rho A^2}\log\left(1+\gamma\right).
\end{equation}
Then, by \eqref{eq:tube:uxgamma}, \eqref{eq:tube:impulse}, and \eqref{eq:tube:p0:MR:bounds}, the necessary and sufficient condition that oscillatory motions occur is that the nonlinear shear modulus, $\widetilde{\mu}$, is uniformly bounded from below as follows,
\begin{equation}\label{eq:tube:shearmod:bound:MR}
\widetilde{\mu}>\frac{p_{0}\rho A^2}{\log\left(1+\gamma\right)}=\alpha\frac{T_{1}(t)-T_{2}(t)}{\log B-\log A}.
\end{equation}
By \eqref{eq:tube:shearmod},
\[
\widetilde{\mu}=\mu_{1}+\mu_{2}\alpha^2=\mu_{1}+\left(\mu-\mu_{1}\right)\alpha^2=\mu\alpha^2+\mu_{1}\left(1-\alpha^2\right).
\]
Hence, \eqref{eq:tube:shearmod:bound:MR} is equivalent to
\begin{equation}\label{eq:tube:mu:bound:MR}
\mu>\frac{p_{0}\rho A^2}{\alpha^2\log\left(1+\gamma\right)}+\mu_{1}\frac{1-\alpha^2}{\alpha^2}.
\end{equation}
Then, the probability distribution of oscillatory motions occurring is
\begin{equation}\label{eq:tube:P1}
P_{1}(p_0)=1-\int_{0}^{\frac{p_{0}\rho A^2}{\alpha^2\log\left(1+\gamma\right)}+\mu_{1}\frac{1-\alpha^2}{\alpha^2}}g(u;\rho_{1},\rho_{2})du,
\end{equation}
where $g(u;\rho_{1},\rho_{2})$ is the Gamma probability density function defined by \eqref{eq:mu:gamma}, and that of non-oscillatory motions is
\begin{equation}\label{eq:tube:P2}
P_{2}(p_0)=1-P_{1}(p_0)=\int_{0}^{\frac{p_{0}\rho A^2}{\alpha^2\log\left(1+\gamma\right)}+\mu_{1}\frac{1-\alpha^2}{\alpha^2}}g(u;\rho_{1},\rho_{2})du.
\end{equation}

\begin{figure}[h!!!t!!!b!!!]
	\begin{center}
		\includegraphics[width=\textwidth]{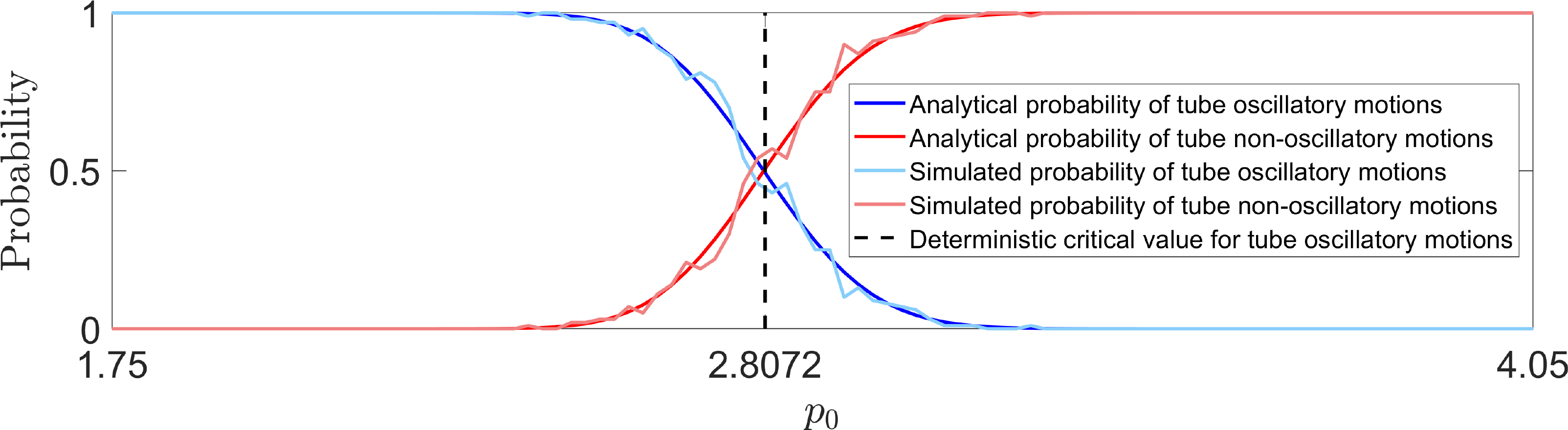}
		\caption{Probability distributions of whether oscillatory motions can occur or not for a cylindrical tube of stochastic Mooney-Rivin material, with $\alpha=1$, $\rho=1$, $A=1$, $\gamma=1$, and the shear modulus, $\mu$, following the Gamma distribution with $\rho_{1}=405$, $\rho_{2}=0.01$. Dark coloured lines represent analytically derived solutions, given by equations \eqref{eq:tube:P1}-\eqref{eq:tube:P2}, whereas the lighter versions represent stochastically generated data. The vertical line at the critical value, $p_{0}=2.8072$, separates the expected regions based only on mean value, $\underline{\mu}=\rho_{1}\rho_{2}=4.05$. The probabilities were calculated from the average of 100 stochastic simulations.}\label{fig:intpdfs-tube}
	\end{center}
	\begin{center}
		\includegraphics[width=0.45\textwidth]{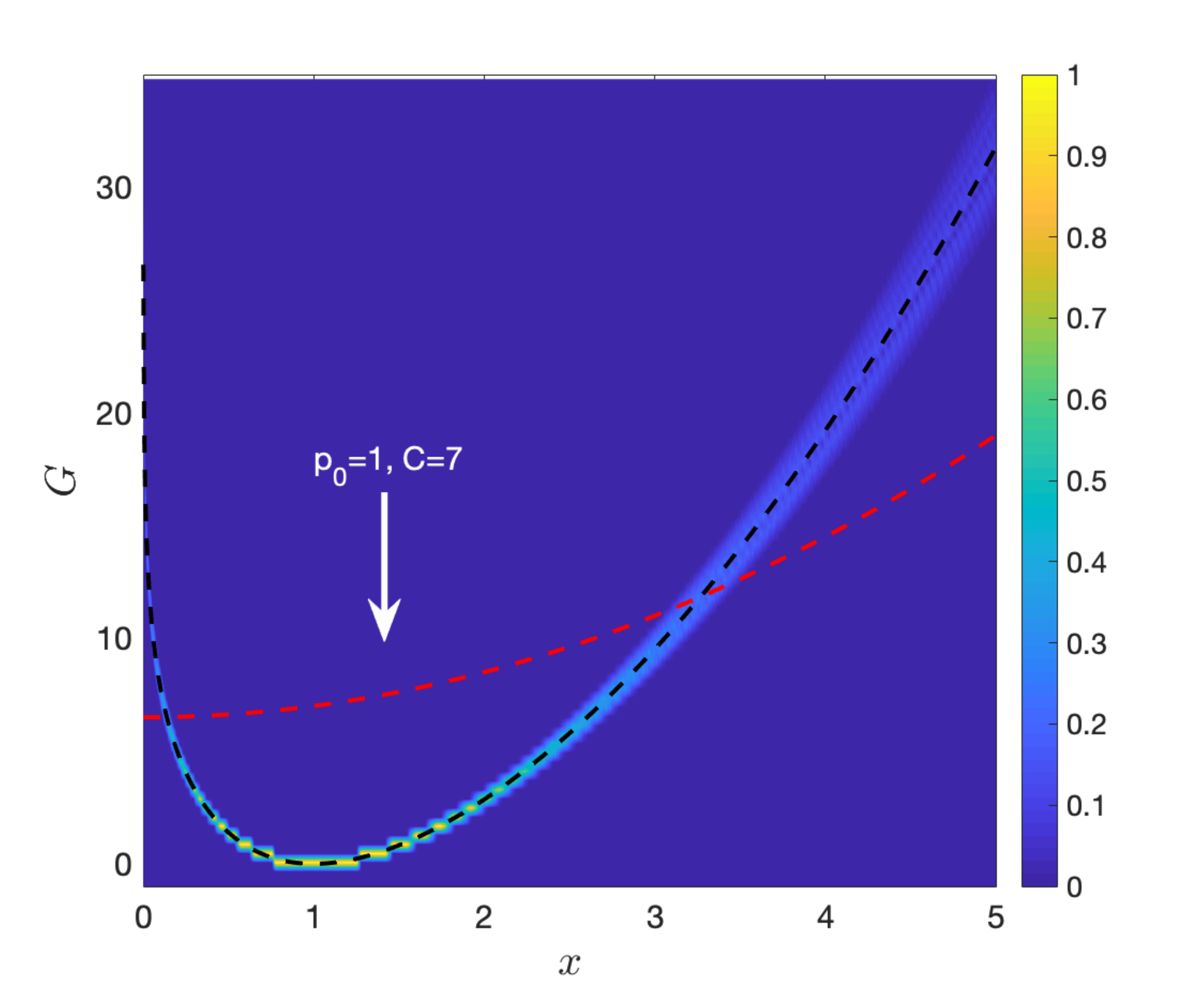}\qquad
		\includegraphics[width=0.45\textwidth]{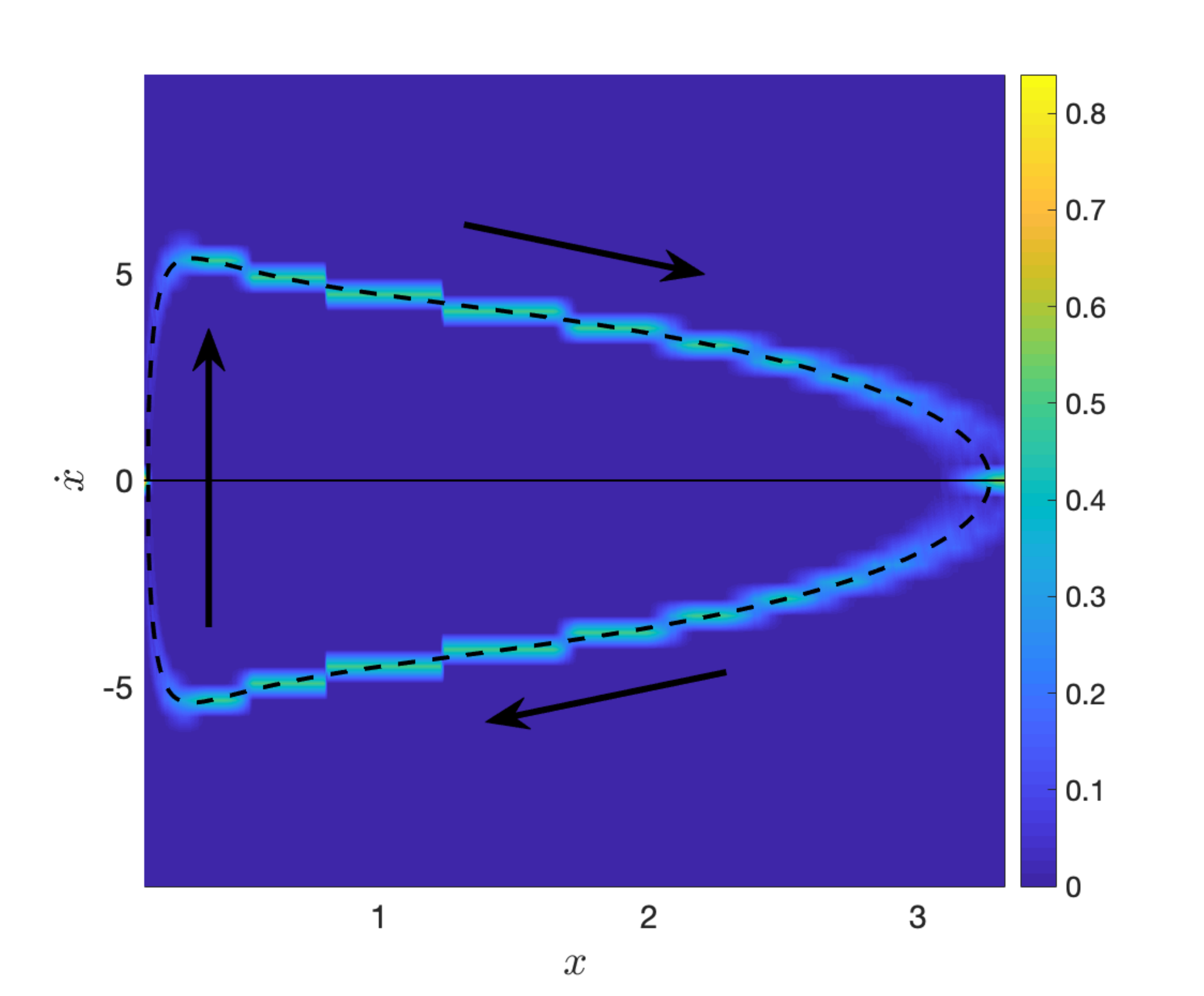}
		\caption{The function $G(x,\gamma)$, defined by \eqref{eq:tube:G:MR}, intersecting the (dashed red) curve $p_{0}\left(x^2-1/\alpha\right)/(2\alpha)+C$ , with $p_{0}=1$ and $C=7$, (left), and the associated velocity, given by \eqref{eq:tube:dotx} (right), for a cylindrical tube of stochastic Mooney-Rivlin material when $\alpha=1$, $\rho=1$, $A=1$, $\gamma=1$, and $\widetilde{\mu}=\mu=\mu_1+\mu_{2}$ is drawn from the Gamma distribution with $\rho_{1}=405$ and $\rho_{2}=0.01$. The dashed black lines correspond to the expected values based only on mean value, $\underline{\mu}=\rho_{1}\rho_{2}=4.05$. Each distribution was calculated from the average of $1000$ stochastic simulations.}\label{fig:stoch-forcedtube}
	\end{center}
\end{figure}

For example, when $\alpha=1$, $\rho=1$, $A=1$, $\gamma=1$, and $\widetilde{\mu}=\mu=\mu_1+\mu_{2}$ satisfies the Gamma distribution with $\rho_{1}=405$ and $\rho_{2}=0.01$, the probability distributions given by \eqref{eq:tube:P1}-\eqref{eq:tube:P2} are shown in Figure~\ref{fig:intpdfs-tube} (blue lines for $P_{1}$ and red lines for $P_{2}$). Specifically, $(0,\underline{\mu})$, where $\underline{\mu}=\rho_{1}\rho_{2}=4.05$ is the mean value of $\mu$, was divided into $100$ steps, then for each value of $p_{0}$, $100$ random values of $\mu$ were numerically generated from the specified Gamma distribution and compared with the inequalities defining the two intervals for values of $p_{0}$. For the deterministic elastic tube, the critical value $p_{0}=\underline{\mu}\log 2\approx 2.8072$ strictly divides the cases of oscillations occurring or not. For the stochastic problem, for the same critical value, there is, by definition, exactly 50\% chance of that the motion is oscillatory, and 50\% chance that is not. To increase the probability of oscillatory motion ($P_{1}\approx 1$), one must apply a sufficiently small impulse, $p_{0}$, below the expected critical point, whereas a non-oscillatory motion is certain to occur ($P_{2}\approx 1$) if $p_{0}$ is sufficiently large. However, the inherent variability in the probabilistic system means that there will also exist events where there is competition between the two cases.

In Figure~\ref{fig:stoch-forcedtube}, we illustrate the stochastic function $G(x,\gamma)$, defined by \eqref{eq:tube:G:MR}, intersecting the curve  $p_{0}\left(x^2-1/\alpha\right)/(2\alpha)+C$, with $p_{0}=1$ and $C=7$, to find the solutions of equation \eqref{eq:tube:GC}, and the associated velocity, given by \eqref{eq:tube:dotx},  assuming that $\alpha=1$, $\rho=1$, $A=1$, $\gamma=1$, and $\mu$ satisfies the Gamma distribution with $\rho_{1}=405$ and $\rho_{2}=0.01$ (see Figure~\ref{fig:mu-gpdf}).

When $C=0$, equation \eqref{eq:tube:p0:MR} can be solved explicitly to find the amplitude
\begin{equation}\label{eq:tube:amp:MR}
x_{1}=\sqrt{\frac{\gamma/\alpha}{\left(1+\gamma\right)\exp\left[-\left(p_{0}\rho A^2\right)/(\widetilde{\mu})\right]-1}}
=\sqrt{\frac{\left(B^2-A^2\right)/\alpha}{B^2\exp\left[-2\alpha\left(P_{1}-P_{2}\right)/\widetilde{\mu}\right]-A^2}}.
\end{equation}
Note that, in the static case, by \eqref{eq:tube:P1P2:x:static} and  \eqref{eq:tube:impulse}, at $x=x_{1}$, the required pressure takes the form
\begin{equation}\label{eq:tube:p0:MR:static}
p_{0}^{(s)}=\frac{\widetilde{\mu}}{\alpha x^2\rho A^2}\frac{\gamma-\frac{\gamma}{\alpha x^2}}{1+\frac{\gamma}{\alpha x^2}}+\frac{\widetilde{\mu}}{\rho A^2}\log\frac{1+\gamma}{1+\frac{\gamma}{\alpha x^2}}.
\end{equation}
Thus, the difference between the applied pressure in the static and dynamic case, given by \eqref{eq:tube:p0:MR:static} and \eqref{eq:tube:p0:MR}, with $C=0$, respectively, is
\begin{equation}\label{eq:tube:p0:MR:differ}
p_{0}^{(s)}-p_{0}=\frac{\widetilde{\mu}}{\alpha x^2\rho A^2}\frac{\gamma-\frac{\gamma}{\alpha x^2}}{1+\frac{\gamma}{\alpha x^2}}.
\end{equation}
Hence, $p_{0}^{(s)}<p_{0}$ if $0<x_{1}<\sqrt{\alpha}$, and $p_{0}^{(s)}>p_{0}$ if $x_{1}>\sqrt{\alpha}$.

- If the tube wall is thin \cite{Knowles:1962,Shahinpoor:1971:SN}, then $0<\gamma\ll 1$ and $\alpha=1$, and \eqref{eq:tube:p0:MR} becomes
\begin{equation}\label{eq:tube:p0:MR:thin}
\frac{p_{0}}{\gamma}=\frac{\mu}{\rho A^2}\left(1-\frac{1}{x^2}\right)-\frac{2\alpha C}{x^2-\frac{1}{\alpha}}.
\end{equation}
Then, the necessary and sufficient condition that oscillatory motions occur is that
\begin{equation}\label{eq:tube:p0:MR:thin:bounds}
-\infty=\lim_{x\to0}\left[\frac{\mu}{\rho A^2}\left(1-\frac{1}{x^2}\right)-\frac{2\alpha C}{x^2-\frac{1}{\alpha}}\right]<\frac{p_{0}}{\gamma}<\lim_{x\to\infty}\left[\frac{\mu}{\rho A^2}\left(1-\frac{1}{x^2}\right)-\frac{2\alpha C}{x^2-\frac{1}{\alpha}}\right]=\frac{\mu}{\rho A^2}.
\end{equation}
Thus, for the motion to be oscillatory, the shear modulus must be bounded from below as follows,
\begin{equation}\label{eq:tube:shearmod:bound:MR:thin}
\mu>\frac{p_{0}}{\gamma}\rho A^2=\frac{2}{\gamma}\left(T_{1}(t)-T_{2}(t)\right).
\end{equation}
Then, the probability distribution of oscillatory motions occurring is
\begin{equation}\label{eq:tube:P1:thin}
P_{1}(p_0/\gamma)=1-\int_{0}^{\frac{p_{0}}{\gamma}\rho A^2}g(u;\rho_{1},\rho_{2})du,
\end{equation}
and that of non-oscillatory motions is
\begin{equation}\label{eq:tube:P2:thin}
P_{2}(p_0/\gamma)=1-P_{1}(p_0/\gamma)=\int_{0}^{\frac{p_{0}}{\gamma}\rho A^2}g(u;\rho_{1},\rho_{2})du.
\end{equation}

\begin{figure}[h!!!t!!!b!!!]
	\begin{center}
		\includegraphics[width=\textwidth]{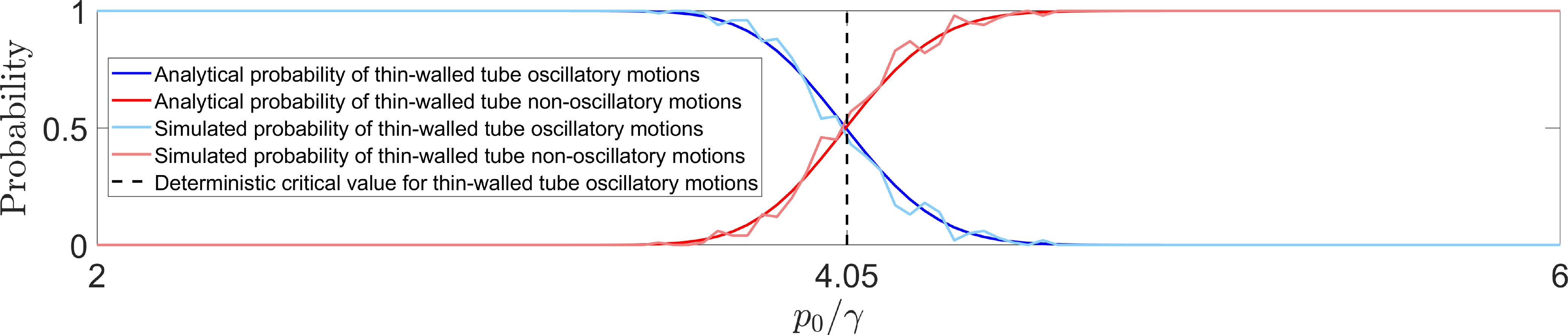}
		\caption{Probability distributions of whether oscillatory motions can occur or not for a thin-walled cylindrical tube of stochastic Mooney-Rivin material, with $\rho=1$, $A=1$, and the shear modulus, $\mu$, following the Gamma distribution with $\rho_{1}=405$, $\rho_{2}=0.01$. Dark coloured lines represent analytically derived solutions, given by equations \eqref{eq:tube:P1}-\eqref{eq:tube:P2}, whereas the lighter versions represent stochastically generated data. The vertical line at the critical value, $p_{0}/\gamma=4.05$, separates the expected regions based only on mean value, $\underline{\mu}=\rho_{1}\rho_{2}=4.05$. The probabilities were calculated from the average of 100 stochastic simulations.}\label{fig:intpdfs-thintube}
	\end{center}
	\begin{center}
		\includegraphics[width=\textwidth]{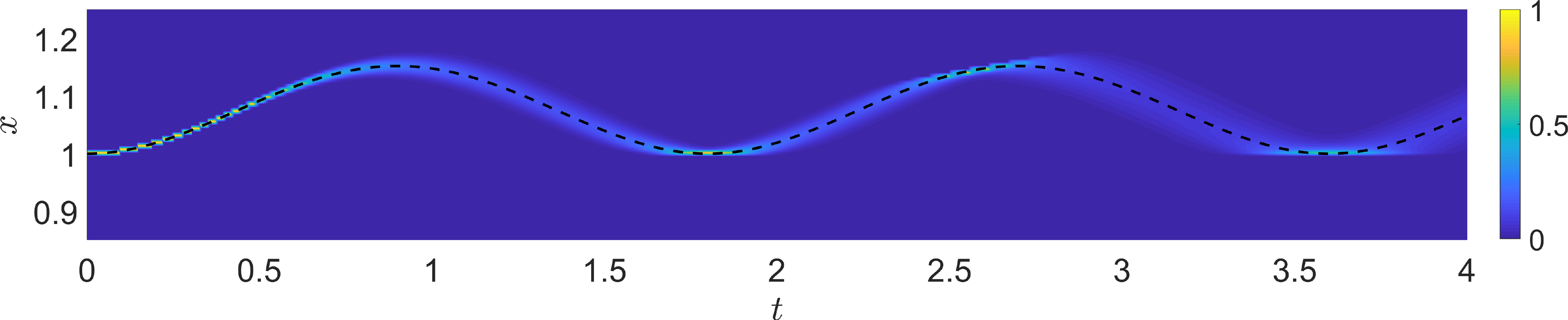}
		\caption{Stochastic solution given by \eqref{eq:tube:x:MR:thin:p0}, with $p_{0}/\gamma=1$, for a thin-walled tube, where $\rho=1$, $A=1$, and $\mu$ is drawn from the Gamma distribution with $\rho_{1}=405$ and $\rho_{2}=0.01$. The dashed black line corresponds to the expected values based only on mean value, $\underline{\mu}=\rho_{1}\rho_{2}=4.05$. The distribution was calculated from the average of $1000$ stochastic simulations.}\label{fig:stochwave-forcedthintube}
	\end{center}
\end{figure}

For $\rho=1$, $A=1$, and $\widetilde{\mu}=\mu=\mu_1+\mu_{2}$ drawn from the Gamma distribution with $\rho_{1}=405$ and $\rho_{2}=0.01$, the probability distributions given by \eqref{eq:tube:P1:thin}-\eqref{eq:tube:P2:thin} are shown in Figure~\ref{fig:intpdfs-thintube} (blue lines for $P_{1}$ and red lines for $P_{2}$). For the deterministic thin-walled tube, the critical value $p_{0}/\gamma=\underline{\mu}=4.05$ strictly separates the cases of oscillations occurring or not. However, in the stochastic case, the two cases compete.

If $C=0$, then setting $x_{0}=1$ and $\dot{x}_{0}=0$, the equation of motion has the explicit solution \cite{Shahinpoor:1971:SN}
\begin{equation}\label{eq:tube:x:MR:thin:p0}
x=\sqrt{\frac{\frac{\mu}{\rho A^2}-\frac{p_{0}}{2\gamma}}{\frac{\mu}{\rho A^2}-\frac{p_{0}}{\gamma}}-\frac{\frac{p_{0}}{2\gamma}}{\frac{\mu}{\rho A^2}-\frac{p_{0}}{\gamma}}\cos\left(2t\sqrt{\frac{\mu}{\rho A^2}-\frac{p_{0}}{\gamma}}\right)}.
\end{equation}
In Figure~\ref{fig:stochwave-forcedthintube}, we illustrate the stochastic solution given by \eqref{eq:tube:x:MR:thin:p0}, with $p_{0}/\gamma=1$, assuming that $\rho=1$, $A=1$, and $\mu$ satisfies the Gamma distribution with hyperparameters $\rho_{1}=405$ and $\rho_{2}=0.01$.

- If the tube wall is infinitely thick \cite{Shahinpoor:1973}, then $\gamma\to\infty$, and assuming that the nonlinear shear modulus, $\widetilde{\mu}$, has a uniform lower bound, \eqref{eq:tube:p0:MR:bounds} becomes
\begin{equation}\label{eq:tube:p0:thickbounds:MR}
-\infty=\lim_{x\to0}\left[\frac{\widetilde{\mu}}{\rho A^2}\log\left(\alpha x^2\right)-\frac{2\alpha C}{x^2-\frac{1}{\alpha}}\right]<p_{0}<\lim_{x\to\infty}\left[\frac{\widetilde{\mu}}{\rho A^2}\log\left(\alpha x^2\right)-\frac{2\alpha C}{x^2-\frac{1}{\alpha}}\right]=\infty.
\end{equation}
Hence, the motion is always oscillatory for any value of the applied impulse.

\section{Quasi-equilibrated radial motion of a stochastic hyperelastic spherical shell}\label{sec:sphere}

Next, we examine the stability and finite amplitude oscillations of a stochastic hyperelastic spherical shell under quasi-equilibrated dynamic radial deformation.

\subsection{Dynamic radial deformation of a spherical shell}

\begin{figure}[htbp]
	\begin{center}
		\includegraphics[width=0.9\textwidth]{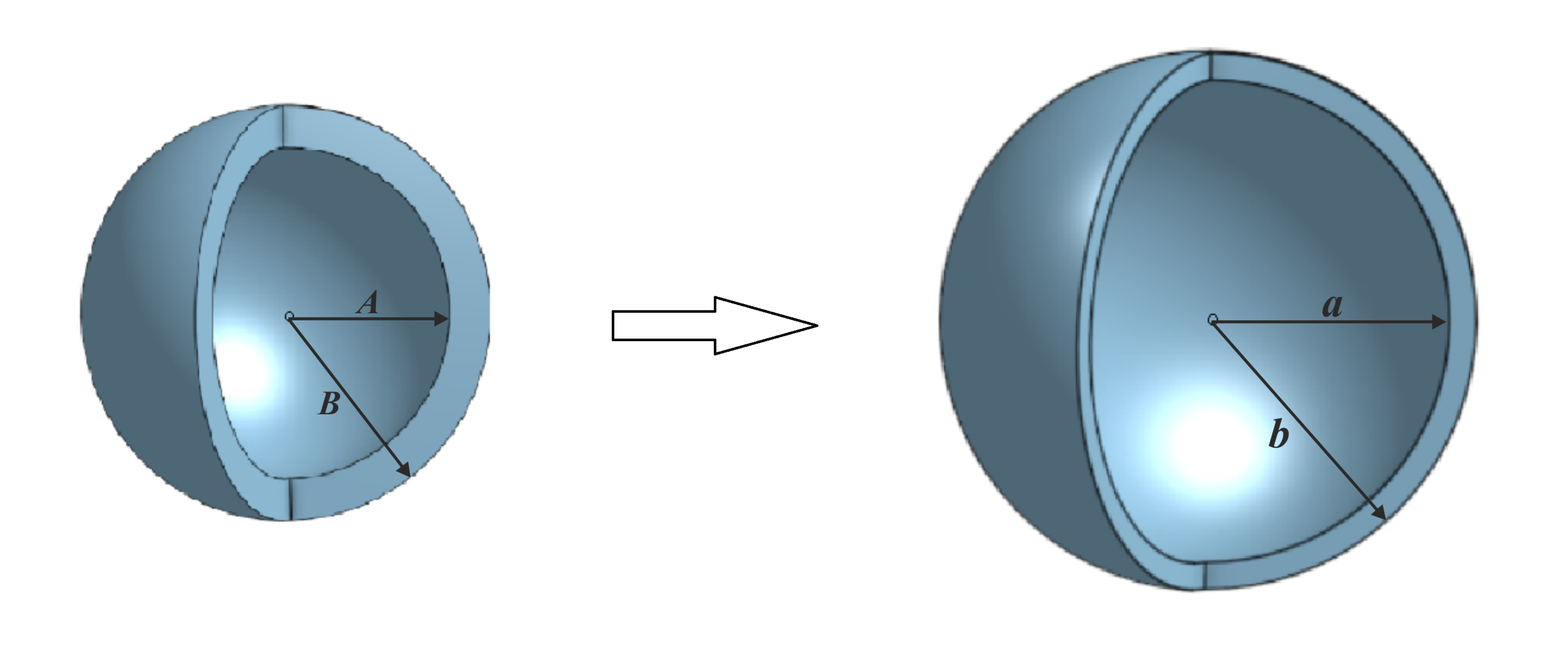}
		\caption{Schematic of inflation of a spherical shell, showing the reference state, with inner radius $A$ and outer radius $B$ (left), and the deformed state, with inner radius $a$ and outer radius $b$ (right), respectively.}\label{fig:shell}
	\end{center}
\end{figure}

For a spherical shell, the radial motion is described by \cite{Balakrishnan:1978:BS,Beatty:2011,Heng:1963:HS,Knowles:1965:KJ} (see Figure~\ref{fig:shell})
\begin{equation}\label{eq:sphere:deform}
r^3=a^3+R^3-A^3,\qquad \theta=\Theta,\qquad \phi=\Phi,
\end{equation}
where $(R,\Theta,\Phi)$ and $(r,\theta,\phi)$ are the spherical polar coordinates in the reference and current configuration, respectively, such that $A\leq R\leq B$, $A$ and $B$ are the inner and outer radii in the undeformed state, and $a=a(t)$  and $b=b(t)=\sqrt[3]{a^3+B^3-A^3}$ are the inner and outer radii at time $t$, respectively.

As for the cylindrical tube, the radial motion \eqref{eq:sphere:deform} of the spherical shell is determined entirely by the inner radius $a$ at time $t$.
By the governing equations \eqref{eq:sphere:deform}, the condition \eqref{eq:ddotx:curl} is valid for $\textbf{x}=(r,\theta,\phi)^{T}$, since
\begin{equation}\label{eq:sphere:curl}
\textbf{0}=\mathrm{curl}\ \ddot{\textbf{x}}=
\left[
\begin{array}{c}
(\partial\ddot{\phi}/\partial\theta)/r-(\partial\ddot{\theta}/\partial\phi)/(r\sin\theta)\\
(\partial\ddot{r}/\partial\phi)/(r\sin\theta)-\partial\ddot{\phi}/\partial r\\
\partial\ddot{\theta}/\partial r-(\partial\ddot{r}/\partial\theta)/r
\end{array}
\right],
\end{equation}
and the acceleration potential, $\xi$, satisfies \eqref{eq:cp}. Hence, this is a quasi-equilibrated motion, such that
\begin{equation}\label{eq:sphere:cp}
-\frac{\partial\xi}{\partial r}=\ddot{r}=\frac{2a\dot{a}^2+a^2\ddot{a}}{r^2}-\frac{2a^4\dot{a}^2}{r^5},
\end{equation}
and integrating \eqref{eq:sphere:cp} gives \cite[p.~217]{TruesdellNoll:2004}
\begin{equation}\label{eq:sphere:xi}
-\xi=-\frac{2a\dot{a}^2+a^2\ddot{a}}{r}+\frac{a^4\dot{a}^2}{2r^4}=-r\ddot{r}-\frac{3}{2}\dot{r}^2.
\end{equation}
For the deformation \eqref{eq:sphere:deform}, the gradient tensor with respect to the polar coordinates $(R,\Theta,\Phi)$ takes the form
\begin{equation}\label{eq:sphere:F}
\textbf{F}=\mathrm{diag}\left(\frac{R^2}{r^2}, \frac{r}{R}, \frac{r}{R}\right),
\end{equation}
the Cauchy-Green tensor is equal to
\begin{equation}\label{eq:sphere:B}
\textbf{B}=\textbf{F}^2=\mathrm{diag}\left(\frac{R^4}{r^4}, \frac{r^2}{R^2}, \frac{r^2}{R^2}\right),
\end{equation}
and the corresponding principal invariants are
\begin{equation}\label{eq:sphere:I123}
\begin{split}
I_{1}=&\mathrm{tr}\ (\textbf{B})=\frac{R^4}{r^4}+2\frac{r^2}{R^2},\\
I_{2}=&\frac{1}{2}\left[\left(\mathrm{tr}\,\textbf{B}\right)^{2}-\mathrm{tr}\left(\textbf{B}^{2}\right)\right]=\frac{r^4}{R^4}+2\frac{R^2}{r^2},\\
I_{3}=&\det\textbf{B}=1.
\end{split}
\end{equation}
Then, the principal components of the equilibrium Cauchy stress at time $t$ are
\begin{equation}\label{eq:sphere:stresses}
\begin{split}
T^{(0)}_{rr}&=-p^{(0)}+\beta_{1}\frac{R^4}{r^4}+\beta_{-1}\frac{r^4}{R^4},\\
T^{(0)}_{\theta\theta}&=T^{(0)}_{rr}+\left(\beta_{1}-\beta_{-1}\frac{r^2}{R^2}\right)\left(\frac{r^2}{R^2}-\frac{R^4}{r^4}\right),\\
T^{(0)}_{\phi\phi}&=T^{(0)}_{\theta\theta},
\end{split}
\end{equation}
where $p^{(0)}$ is the Lagrangian multiplier for the incompressibility constraint ($I_{3}=1$), and
\begin{equation}\label{eq:sphere:betas}
\beta_{1}=2\frac{\partial W}{\partial I_{1}},\qquad \beta_{-1}=-2\frac{\partial W}{\partial I_{2}},
\end{equation}
with $I_{1}$ and $I_{2}$ given by \eqref{eq:sphere:I123}.

As the stress components depend only on the radius $r$, the system of equilibrium equations reduces to
\begin{equation}\label{eq:sphere:equilibrium}
\frac{\partial T^{(0)}_{rr}}{\partial r}=2\frac{T^{(0)}_{\theta\theta}-T^{(0)}_{rr}}{r}.
\end{equation}
Hence, by \eqref{eq:sphere:stresses} and \eqref{eq:sphere:equilibrium}, the radial Cauchy stress for the equilibrium state at $t$ is equal to
\begin{equation}\label{eq:sphere:T0rr}
T^{(0)}_{rr}(r,t)=2\int\left(\beta_{1}-\beta_{-1}\frac{r^2}{R^2}\right)\left(\frac{r^2}{R^2}-\frac{R^4}{r^4}\right)\frac{dr}{r}+\psi(t),
\end{equation}
where $\psi=\psi(t)$ is an arbitrary function of time. Substitution of \eqref{eq:sphere:xi} and \eqref{eq:sphere:T0rr} into \eqref{eq:T} gives the following principal Cauchy stresses at time $t$,
\begin{equation}\label{eq:sphere:Trr}
\begin{split}
T_{rr}(r,t)&=-\rho\left(\frac{a^2\ddot{a}+2a\dot{a}^2}{r}-\frac{a^4\dot{a}^2}{2r^4}\right)+2\int\left(\beta_{1}-\beta_{-1}\frac{r^2}{R^2}\right)\left(\frac{r^2}{R^2}-\frac{R^4}{r^4}\right)\frac{dr}{r}+\psi(t),\\
T_{\theta\theta}(r,t)&=T_{rr}(r,t)+\left(\beta_{1}-\beta_{-1}\frac{r^2}{R^2}\right)\left(\frac{r^2}{R^2}-\frac{R^4}{r^4}\right),\\
T_{\phi\phi}(r,t)&=T_{\theta\theta}(r,t).
\end{split}
\end{equation}
In \eqref{eq:sphere:Trr}, the function $\beta_{1}-\beta_{-1}\left(r^2/R^2\right)$ can be regarded as the following nonlinear shear modulus \cite{Beatty:2011,Mihai:2017:MG}
\begin{equation}\label{eq:sphere:shearmod}
\widetilde{\mu}=\beta_{1}-\beta_{-1}\frac{r^2}{R^2},
\end{equation}
corresponding to the combined deformation of infinitesimal shear superposed on finite axial stretch, defined by \eqref{eq:stretchshear}, with the shear parameter satisfying $k\to0$ and the stretch parameter $\alpha=r/R$. This modulus is positive if the BE inequalities \eqref{eq:BE} hold \cite{Mihai:2017:MG}. In this case, the integrand in \eqref{eq:sphere:Trr} is negative for $0<r^2/R^2<1$ (i.e., when $0<a^2/A^2<1$) and positive for $r^2/R^2>1$ (i.e., when $a^2/A^2>1$).

When $R^{2}/r^{2}\to 1$, the nonlinear elastic modulus given by \eqref{eq:sphere:shearmod} converges to the shear modulus from linear elasticity,
\begin{equation}\label{eq:sphere:shearmod:lin}
\mu=\lim_{R^{2}/r^{2}\to 1}\widetilde{\mu}.
\end{equation}
In this case, the stress components given by \eqref{eq:sphere:Trr} are equal.

For the spherical shell deforming by \eqref{eq:sphere:deform}, we set the inner and outer radial pressures acting on the curvilinear surfaces, $r=a(t)$ and $r=b(t)$ at time $t$, as $T_{1}(t)$ and $T_{2}(t)$, respectively \cite[pp.~217-219]{TruesdellNoll:2004}.  Then, evaluating $T_{1}(t)=-T_{rr}(a,t)$ and $T_{2}(t)=-T_{rr}(b,t)$, using \eqref{eq:sphere:Trr}, with $r=a$ and $r=b$, respectively, and subtracting the results, gives
\begin{equation}\label{eq:sphere:P1P2:r}
\begin{split}
T_{1}(t)-T_{2}(t)&=\rho\left[\left(a^2\ddot{a}+2a\dot{a}^2\right)\left(\frac{1}{a}-\frac{1}{b}\right)-\frac{a^4\dot{a}^2}{2}\left(\frac{1}{a^4}-\frac{1}{b^4}\right)\right]+2\int_{a}^{b}\widetilde{\mu}\left(\frac{r^2}{R^2}-\frac{R^4}{r^4}\right)\frac{dr}{r}\\
&=\rho\left[\left(a\ddot{a}+2\dot{a}^2\right)\left(1-\frac{a}{b}\right)-\frac{\dot{a}^2}{2}\left(1-\frac{a^4}{b^4}\right)\right]+2\int_{a}^{b}\widetilde{\mu}\left(\frac{r^2}{R^2}-\frac{R^4}{r^4}\right)\frac{dr}{r}\\
&=\rho A^2\left[\left(\frac{a}{A}\frac{\ddot{a}}A+2\frac{\dot{a}^2}{A^2}\right)\left(1-\frac{a}{b}\right)-\frac{\dot{a}^2}{2A^2}\left(1-\frac{a^4}{b^4}\right)\right]+2\int_{a}^{b}\widetilde{\mu}\left(\frac{r^2}{R^2}-\frac{R^4}{r^4}\right)\frac{dr}{r}
\end{split}
\end{equation}
Setting the notation
\begin{equation}\label{eq:sphere:uxgamma}
u=\frac{r^3}{R^3}=\frac{r^3}{r^3-a^3+A^3},\qquad
x=\frac{a}{A},\qquad
\gamma=\frac{B^3}{A^3}-1,
\end{equation}
we can rewrite
\[
\begin{split}
\left(\frac{a}{A}\frac{\ddot{a}}A+2\frac{\dot{a}^2}{A^2}\right)\left(1-\frac{a}{b}\right)&-\frac{\dot{a}^2}{2A^2}\left(1-\frac{a^4}{b^4}\right)\\
&=\left(\ddot{x}x+2\dot{x}^2\right)\left[1-\left(1+\frac{\gamma}{x^3}\right)^{-1/3}\right]-\frac{\dot{x}^2}{2}\left[1-\left(1+\frac{\gamma}{x^3}\right)^{-4/3}\right]\\
&=\left(\ddot{x}x+\frac{3}{2}\dot{x}^2\right)\left[1-\left(1+\frac{\gamma}{x^3}\right)^{-1/3}\right]-\frac{\dot{x}^2}{2}\frac{\gamma}{x^3}\left(1+\frac{\gamma}{x^3}\right)^{-4/3}\\
&=\frac{1}{2x^2}\frac{d}{dx}\left\{\dot{x}^2x^3\left[1-\left(1+\frac{\gamma}{x^3}\right)^{-1/3}\right]\right\}
\end{split}
\]
and
\[
\begin{split}
\int_{a}^{b}\widetilde{\mu}\left(\frac{r^2}{R^2}-\frac{R^4}{r^4}\right)\frac{dr}{r}
&=\int_{a}^{b}\widetilde{\mu}\left[{\left(\frac{r^3}{r^3-a^3+A^3}\right)^{2/3}}-\left(\frac{r^3-a^3+A^3}{r^3}\right)^{4/3}\right]\frac{dr}{r}\\
&=\frac{1}{3}\int_{\frac{x^3+\gamma}{1+\gamma}}^{x^3}\widetilde{\mu}\frac{1+u}{u^{7/3}}du.
\end{split}
\]
Hence, \eqref{eq:sphere:P1P2:r} can be written equivalently as follows,
\begin{equation}\label{eq:sphere:P1P2:x}
2x^2\frac{T_{1}(t)-T_{2}(t)}{\rho A^2}
=\frac{d}{dx}\left\{\dot{x}^2x^3\left[1-\left(1+\frac{\gamma}{x^3}\right)^{-1/3}\right]\right\}
+\frac{4x^2}{3\rho A^2}\int_{\frac{x^3+\gamma}{1+\gamma}}^{x^3}\widetilde{\mu}\frac{1+u}{u^{7/3}}du.
\end{equation}
Note that, when the BE inequalities \eqref{eq:BE} hold, $\widetilde{\mu}>0$, and the integral in \eqref{eq:sphere:P1P2:x} is negative if $0<x<1$ and positive if $x>1$.

In the static case, \eqref{eq:sphere:P1P2:r} reduces to
\begin{equation}\label{eq:sphere:P1P2:r:static}
	T_{1}(t)-T_{2}(t)=\int_{a}^{b}\widetilde{\mu}\left(\frac{r^2}{R^2}-\frac{R^4}{r^4}\right)\frac{dr}{r},
\end{equation}
and \eqref{eq:sphere:P1P2:x} becomes
\begin{equation}\label{eq:sphere:P1P2:x:static}
	2\frac{T_{1}(t)-T_{2}(t)}{\rho A^2}=\frac{4}{3\rho A^2}\int_{\frac{x^3+\gamma}{1+\gamma}}^{x^3}\widetilde{\mu}\frac{1+u}{u^{7/3}}du.
\end{equation}

For the dynamic spherical shell, we set
\begin{equation}\label{eq:sphere:H}
H(x,\gamma)=\frac{4}{3\rho A^2}\int_{1}^{x}\left(\zeta^2\int_{\frac{\zeta^3+\gamma}{1+\gamma}}^{\zeta^3}\widetilde{\mu}\frac{1+u}{u^{7/3}}du\right)d\zeta,
\end{equation}
and obtain that $H(x,\gamma)$ is monotonically decreasing when $0<x<1$ and increasing when $x>1$.

We also set a pressure impulse that is constant in time,
\begin{equation}\label{eq:sphere:impulse}
2\frac{T_{1}(t)-T_{2}(t)}{\rho A^2}=\left\{
\begin{array}{cc}
0 & \mbox{if}\ t\leq0,\\
p_{0} & \mbox{if}\ t>0.
\end{array}
\right.
\end{equation}
Then, integrating \eqref{eq:sphere:P1P2:x} once gives
\begin{equation}\label{eq:sphere:ode}
\dot{x}^2x^3\left[1-\left(1+\frac{\gamma}{x^3}\right)^{-1/3}\right]+H(x,\gamma)=\frac{p_{0}}{3}\left(x^3-1\right)+C,
\end{equation}
with $H(x,\gamma)$ defined by \eqref{eq:sphere:H}, and
\begin{equation}\label{eq:sphere:C}
C=\dot{x}_{0}^2x_{0}^3\left[1-\left(1+\frac{\gamma}{x^3}\right)^{-1/3}\right]+H(x_{0},\gamma)-\frac{p_{0}}{3}\left(x_{0}^3-1\right),
\end{equation}
where $x(0)=x_{0}$ and $\dot{x}(0)=\dot{x}_{0}$ are the initial conditions. From \eqref{eq:sphere:ode}, we obtain
\begin{equation}\label{eq:sphere:dotx:forced}
\dot{x}=\pm\sqrt{\frac{\frac{p_{0}}{3}\left(x^3-1\right)+C-H(x,\gamma)}{x^3\left[1-\left(1+\frac{\gamma}{x^3}\right)^{-1/3}\right]}}.
\end{equation}
The analogy with the motion of a point mass in a potential still holds with appropriate modification. 
Hence, oscillatory motion of the spherical shell occurs if and only if the following equation,
\begin{equation}\label{eq:sphere:HC}
H(x,\gamma)=\frac{p_{0}}{3}\left(x^3-1\right)+C,
\end{equation}
has exactly two distinct solutions, representing the amplitudes of the oscillation, $x=x_{1}$ and $x=x_{2}$, such that $0<x_{1}<x_{2}<\infty$. In this case, the minimum and maximum radii of the inner surface in the oscillation are given by $x_{1}A$ and $x_{2}A$, respectively, and the period of oscillation is equal to
\begin{equation}\label{eq:sphere:T:}
T=2\left|\int_{x_{1}}^{x_{2}}\frac{dx}{\dot{x}}\right|=2\left|\int_{x_{1}}^{x_{2}}\sqrt{\frac{x^3\left[1-\left(1+\frac{\gamma}{x^3}\right)^{-1/3}\right]}{\frac{p_{0}}{3}\left(x^3-1\right)+C-H(x,\gamma)}}dx\right|.
\end{equation}
Note that the amplitude and the period of the oscillation are random variables characterised by probability distributions.

\subsection{Radial oscillations of a spherical shell of stochastic neo-Hookean material}

For a spherical shell of stochastic neo-Hookean material, with $\mu_{1}=\mu>0$ and $\mu_{2}=0$ in \eqref{eq:W:stoch}, evaluating the integral in \eqref{eq:sphere:H} gives (see Appendix~\ref{sec:append} for a detailed derivation)
\begin{equation}\label{eq:sphere:H:NH}
H(x,\gamma)=\frac{\mu}{\rho A^2}\left(x^3-1\right)\left[\frac{2x^3-1}{x^3+x^2+x}
-\frac{2\frac{x^3+\gamma}{1+\gamma}-1}{\frac{x^3+\gamma}{1+\gamma}+\left(\frac{x^3+\gamma}{1+\gamma}\right)^{2/3}+\left(\frac{x^3+\gamma}{1+\gamma}\right)^{1/3}}\right].
\end{equation}
Assuming that the nonlinear shear modulus, $\mu$, is uniformly bounded from below, i.e.,
\begin{equation}\label{eq:sphere:shearnod:bound}
\mu>\eta,
\end{equation}
for some constant $\eta>0$, it follows that
\begin{equation}\label{eq:sphere:H:infty}
\lim_{x\to0}H(x,\gamma)=\lim_{x\to\infty}H(x,\gamma)=\infty.
\end{equation}

\paragraph{(i)}When $p_{0}=0$ and $C>0$, equation \eqref{eq:sphere:HC} has exactly two solutions, $x=x_{1}$ and $x=x_{2}$, satisfying $0<x_{1}<1<x_{2}<\infty$, for any positive constant $C$. In this case, it should be noted that, by \eqref{eq:sphere:Trr}, if $T_{rr}(r,t)=0$ at $r=a$ and $r=b$, so that $T_{1}(t)=T_{2}(t)=0$, then, $T_{\theta\theta}(r,t)=T_{\phi\phi}(r,t)\neq 0$ at $r=a$ and $r=b$, unless $r^3/R^3\to1$. Thus, the oscillations cannot be considered as `free' in general.

\begin{figure}[htbp]
	\begin{center}
		\includegraphics[width=0.45\textwidth]{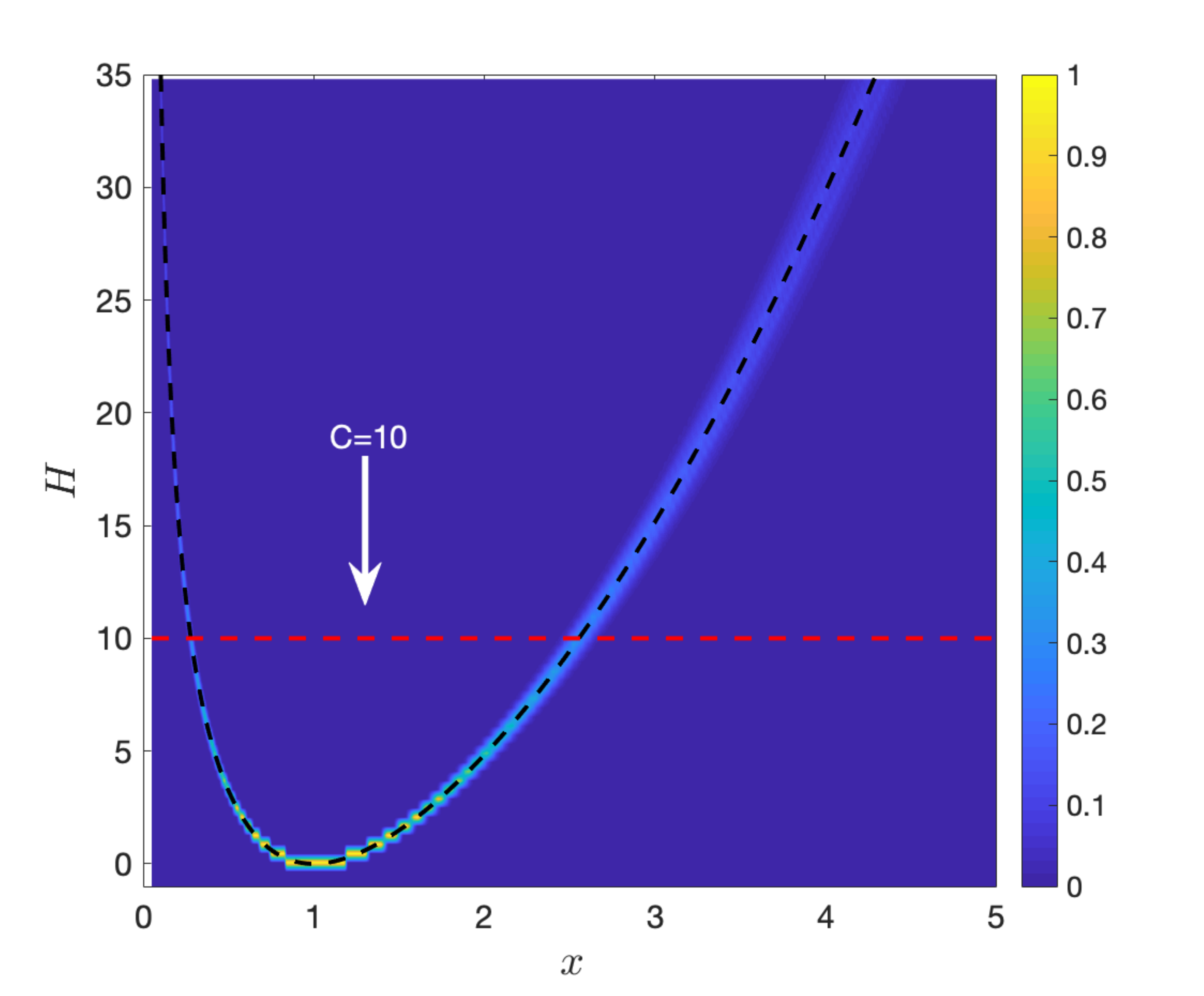}\qquad
		\includegraphics[width=0.45\textwidth]{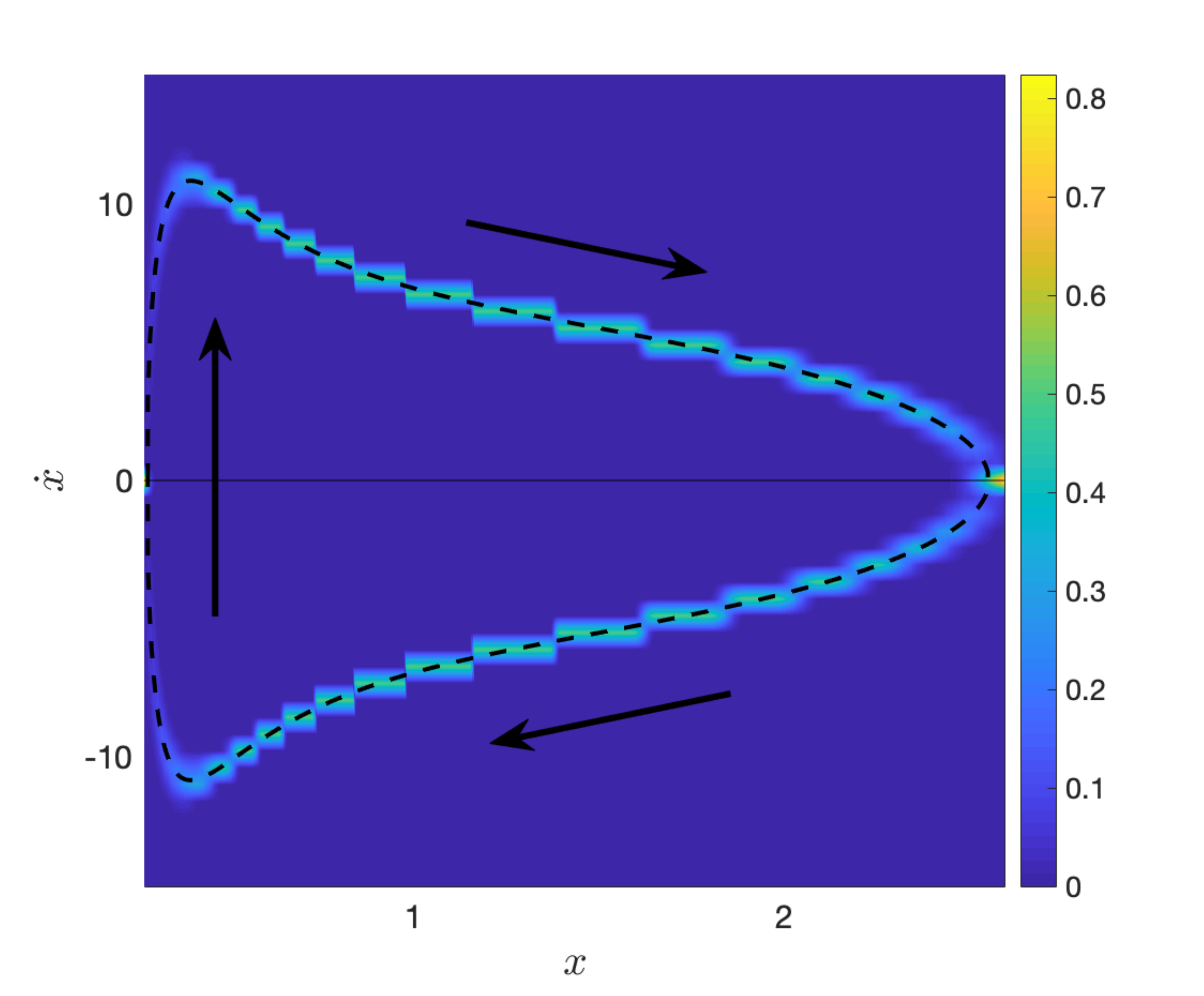}
		\caption{The function $H(x,\gamma)$, defined by \eqref{eq:sphere:H:NH}, intersecting the (dashed red) line $C=10$, when $p_{0}=0$ (left), and the associated velocity, given by \eqref{eq:sphere:dotx:forced} (right), for the spherical shell of stochastic neo-Hookean material, where $\rho=1$, $A=1$, $\gamma=1$, and $\mu$ is drawn from the Gamma distribution with $\rho_{1}=405$ and $\rho_{2}=0.01$. The dashed black lines correspond to the expected values based only on mean value, $\underline{\mu}=\rho_{1}\rho_{2}=4.05$. Each distribution was calculated from the average of $1000$ stochastic simulations.}\label{fig:stoch-freeshell}
	\end{center}
\end{figure}

In Figure~\ref{fig:stoch-freeshell}, we show the stochastic function $H(x,\gamma)$, defined by \eqref{eq:sphere:H:NH}, intersecting the line $C=10$, to solve equation \eqref{eq:sphere:HC} when $p_{0}=0$, and the  associated velocity, given by \eqref{eq:sphere:dotx:forced}, assuming that $\rho=1$, $A=1$, $\gamma=1$, and $\mu$ follows the Gamma distribution with hyperparameters $\rho_{1}=405$ and $\rho_{2}=0.01$ (see Figure~\ref{fig:mu-gpdf}).

\paragraph{(ii)}When $p_{0}\neq0$ and $C\geq 0$, substitution of \eqref{eq:sphere:H:NH} in \eqref{eq:sphere:HC} gives
\begin{equation}\label{eq:sphere:p0:NH}
p_{0}=\frac{3\mu}{\rho A^2}\left[\frac{2x^3-1}{x^3+x^2+x}
-\frac{2\frac{x^3+\gamma}{1+\gamma}-1}{\frac{x^3+\gamma}{1+\gamma}+\left(\frac{x^3+\gamma}{1+\gamma}\right)^{2/3}+\left(\frac{x^3+\gamma}{1+\gamma}\right)^{1/3}}\right]-\frac{3C}{x^3-1}.
\end{equation}
As the right-hand side of \eqref{eq:sphere:p0:NH} is function of $x$ that monotonically increases from $-\infty$ as $x\to 0$ to $\infty$ as $x\to\infty$, the motion is oscillatory for all values of the given pressure difference.

In the static case, by \eqref{eq:sphere:P1P2:x:static} and \eqref{eq:sphere:impulse}, the applied pressure takes the form
\begin{equation}\label{eq:sphere:p0:NH:static}
p_{0}^{(s)}=\frac{\mu}{\rho A^2}\left[\left(\frac{1+\gamma}{x^3+\gamma}\right)^{4/3}+4\left(\frac{1+\gamma}{x^3+\gamma}\right)^{1/3}-\frac{1}{x^4}-\frac{4}{x}\right].
\end{equation}

\begin{figure}[htbp]
	\begin{center}
		\includegraphics[width=0.45\textwidth]{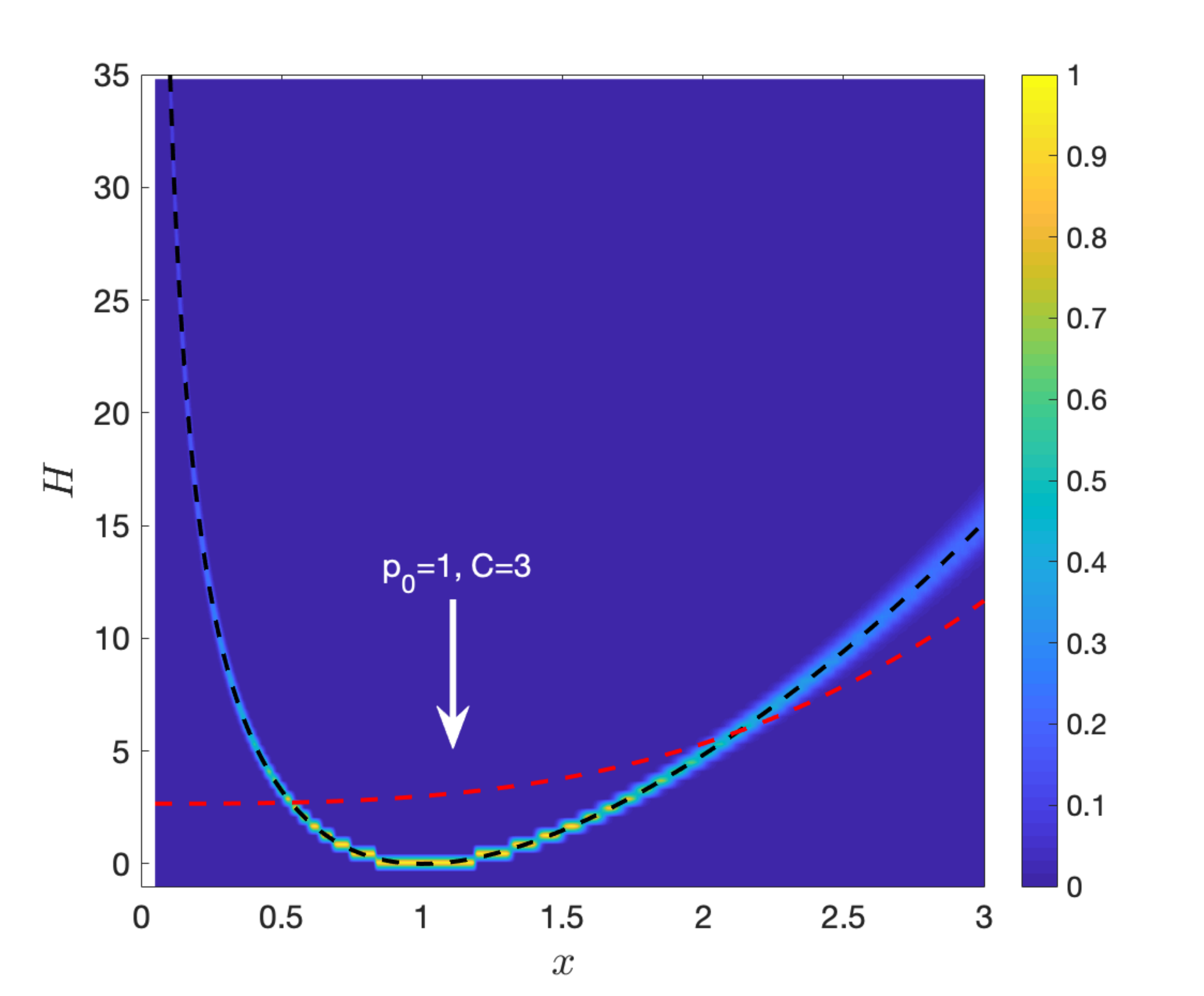}\qquad
		\includegraphics[width=0.45\textwidth]{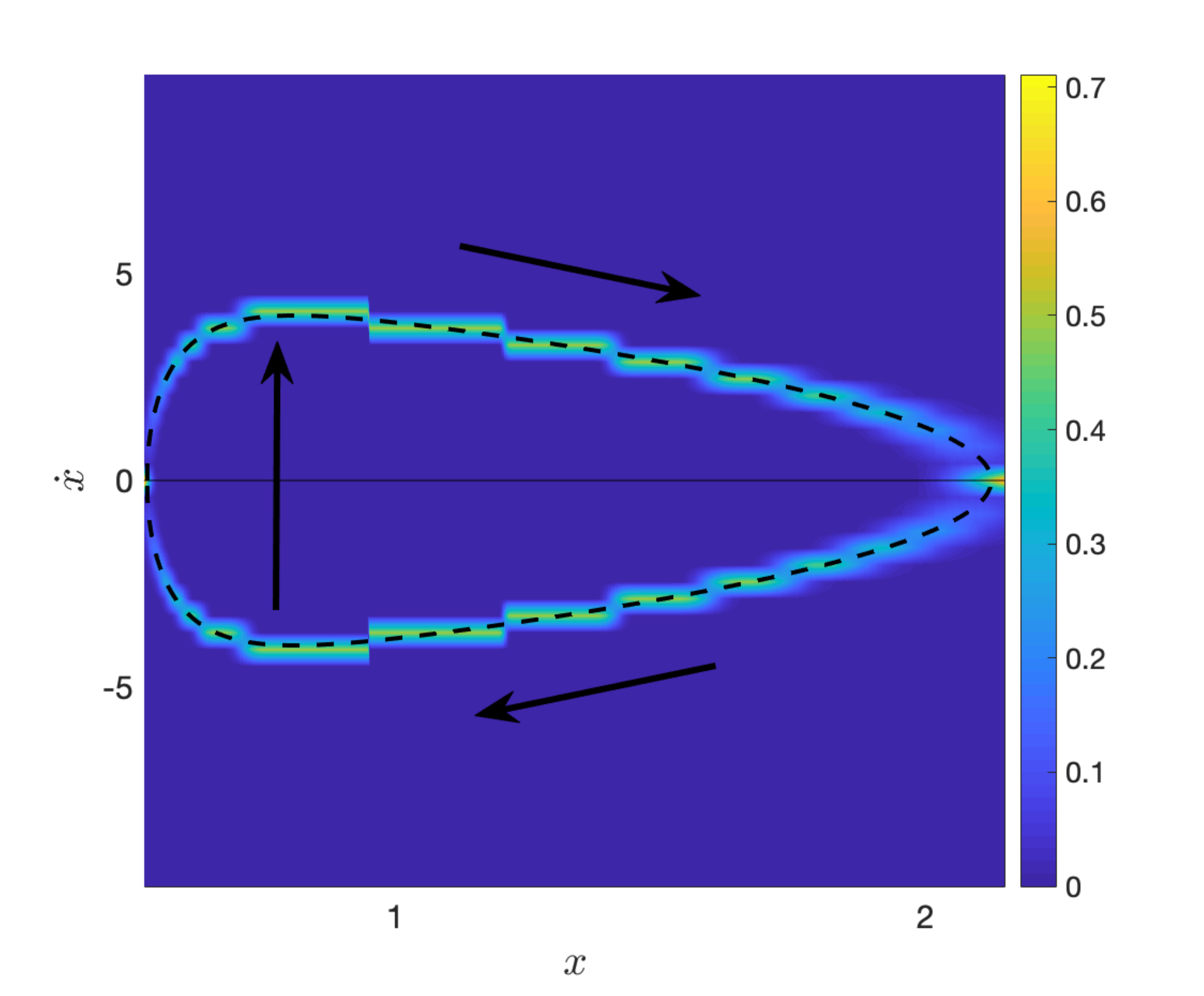}
		\caption{The function $H(x,\gamma)$, defined by \eqref{eq:sphere:H:NH}, intersecting the (dashed red) curve $p_{0}\left(x^3-1\right)/3+C$, with $p_{0}=1$ and $C=3$ (left), and the associated velocity, given by \eqref{eq:sphere:dotx:forced} (right),  for the spherical shell of stochastic neo-Hookean material, where $\rho=1$, $A=1$, $\gamma=1$, and $\mu$ is drawn from the Gamma distribution with $\rho_{1}=405$ and $\rho_{2}=0.01$. The dashed black lines correspond to the expected values based only on mean value, $\underline{\mu}=\rho_{1}\rho_{2}=4.05$. Each distribution was calculated from the average of $1000$ stochastic simulations.}\label{fig:stoch-forcedshell}
	\end{center}
\end{figure}

In Figure~\ref{fig:stoch-forcedshell}, we represent the stochastic function $H(x,\gamma)$, defined by \eqref{eq:sphere:H:NH}, intersecting the curve $p_{0}\left(x^3-1\right)/3+C$, with $p_{0}=1$ and $C=3$,  to obtain the solutions of equation \eqref{eq:sphere:HC}, and the  associated velocity, given by \eqref{eq:sphere:dotx:forced}, assuming that $\rho=1$, $A=1$, $\gamma=1$, and $\mu$ follows the Gamma distribution with $\rho_{1}=405$ and $\rho_{2}=0.01$ (see Figure~\ref{fig:mu-gpdf}).

\begin{figure}[htbp]
	\begin{center}
		\includegraphics[width=\textwidth]{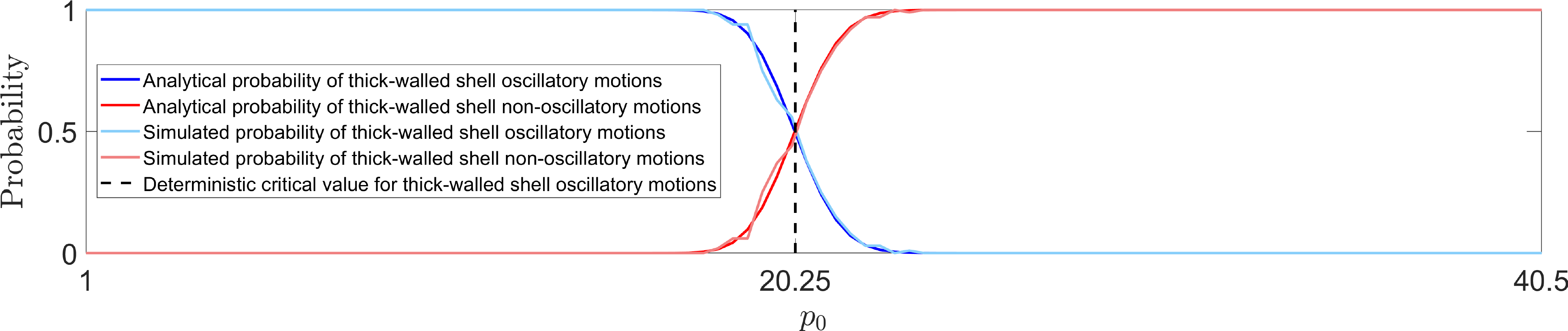}
		\caption{Probability distributions of whether oscillatory motions can occur or not for an infinitely thick-walled spherical shell of stochastic neo-Hookean material, with $\rho=1$, $A=1$, and the shear modulus, $\mu$, following the Gamma distribution with $\rho_{1}=405$, $\rho_{2}=0.01$. Dark coloured lines represent analytically derived solutions, given by equations \eqref{eq:sphere:P1:thick}-\eqref{eq:sphere:P2:thick}, whereas the lighter versions represent stochastically generated data. The vertical line at the critical value, $p_{0}=20.25$, separates the expected regions based only on mean value, $\underline{\mu}=\rho_{1}\rho_{2}=4.05$. The probabilities were calculated from the average of 100 stochastic simulations.}\label{fig:intpdfs-thickshell}
	\end{center}
\end{figure}

- If the spherical shell has an infinitely thick wall \cite{Balakrishnan:1978:BS,Knowles:1965:KJ}, then $\gamma\to\infty$, and the necessary and sufficient condition for  the motion to be oscillatory becomes
\begin{equation*}\label{eq:sphere:p0:limbounds:NH:thick}
\lim_{x\to0}\left[\frac{3\mu}{\rho A^2}\left(\frac{2x^3-1}{x^3+x^2+x}-\frac{1}{3}\right)-\frac{3C}{x^3-1}\right]<p_{0}<\lim_{x\to\infty}\left[\frac{3\mu}{\rho A^2}\left(\frac{2x^3-1}{x^3+x^2+x}-\frac{1}{3}\right)-\frac{3C}{x^3-1}\right],
\end{equation*}
that is
\begin{equation}\label{eq:sphere:p0:bounds:NH:thick}
-\infty<p_{0}<\frac{5\mu}{\rho A^2}.
\end{equation}
Thus, for the oscillations to occur, the shear modulus must satisfy \cite{Knowles:1965:KJ}
\begin{equation}\label{eq:sphere:shearmod:bound:NH:thick}
\mu>p_{0}\frac{\rho A^2}{5}=\frac{2}{5}\left(T_{1}(t)-T_{2}(t)\right).
\end{equation}
Then, the probability distribution of oscillatory motions occurring is
\begin{equation}\label{eq:sphere:P1:thick}
P_{1}(p_0)=1-\int_{0}^{p_{0}\frac{\rho A^2}{5}}g(u;\rho_{1},\rho_{2})du,
\end{equation}
and that of non-oscillatory motions is
\begin{equation}\label{eq:sphere:P2:thick}
P_{2}(p_0)=1-P_{1}(p_0)=\int_{0}^{p_{0}\frac{\rho A^2}{5}}g(u;\rho_{1},\rho_{2})du.
\end{equation}

For $\rho=1$, $A=1$, and $\widetilde{\mu}=\mu=\mu_1+\mu_{2}$ drawn from the Gamma distribution with $\rho_{1}=405$ and $\rho_{2}=0.01$, the probability distributions given by \eqref{eq:sphere:P1:thick}-\eqref{eq:sphere:P2:thick} are shown in Figure~\ref{fig:intpdfs-thickshell} (blue lines for $P_{1}$ and red lines for $P_{2}$). For the deterministic thin-walled tube, the critical value $p_{0}=5\underline{\mu}=20.25$ strictly separates the cases of oscillations occurring or not. However, in the stochastic case, there is competition between the two cases.

\begin{figure}[htbp]
	\begin{center}
		\includegraphics[width=\textwidth]{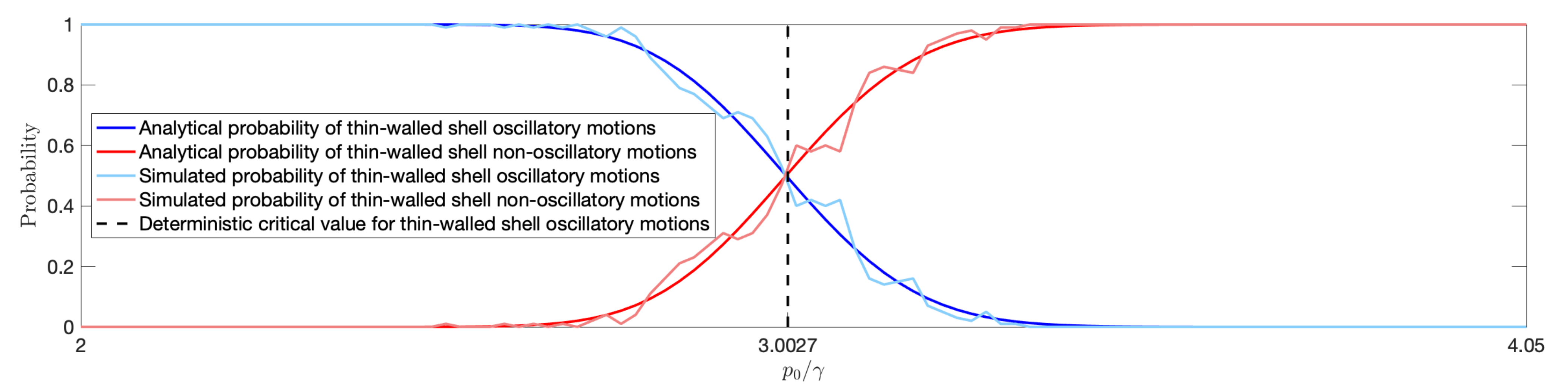}
		\caption{Probability distributions of whether oscillatory motions can occur or not for a thin-walled spherical shell of stochastic neo-Hookean material, with $\rho=1$, $A=1$, and the shear modulus, $\mu$, following the Gamma distribution with $\rho_{1}=405$, $\rho_{2}=0.01$. Dark coloured lines represent analytically derived solutions, given by equations \eqref{eq:sphere:P1:thin}-\eqref{eq:sphere:P2:thin}, whereas the lighter versions represent stochastically generated data. The vertical line at the critical value, $p_{0}/\gamma=3.0027$, separates the expected regions based only on mean value, $\underline{\mu}=\rho_{1}\rho_{2}=4.05$. The probabilities were calculated from the average of 100 stochastic simulations.}\label{fig:intpdfs-thinshell}
	\end{center}
\end{figure}

- If the spherical shell wall is thin \cite{Beatty:2011,Verron:1999:VKDR,Wang:1965}, then $0<\gamma\ll 1$, and setting $C=0$ for example, the necessary and sufficient condition for the oscillatory motions to occur becomes
\begin{equation}\label{eq:sphere:p0:bounds:NH:thin}
-\infty=\lim_{x\to0}\frac{\mu}{\rho A^2}\frac{\left(x+1\right)\left(2x^4-x^2-1\right)}{x^3\left(x^3+x^2+x\right)}<\frac{p_{0}}{\gamma}<\sup_{0<x<\infty}\frac{\mu}{\rho A^2}\frac{\left(x+1\right)\left(2x^4-x^2-1\right)}{x^3\left(x^3+x^2+x\right)}\approx 0.7414\frac{\mu}{\rho A^2},
\end{equation}
where ``$\sup$'' denotes supremum. Hence, for the motion to be oscillatory, the shear modulus must be uniformly bounded from below as follows,
\begin{equation}\label{eq:sphere:shearmod:bound:NH:thin}
\mu>\frac{p_{0}}{\gamma}\frac{\rho A^2}{0.7414}\approx\frac{2.7}{\gamma}\left(T_{1}(t)-T_{2}(t)\right).
\end{equation}
Then, the probability distribution of oscillatory motions occurring is
\begin{equation}\label{eq:sphere:P1:thin}
P_{1}(p_0/\gamma)=1-\int_{0}^{\frac{p_{0}}{\gamma}\frac{\rho A^2}{0.7414}}g(u;\rho_{1},\rho_{2})du,
\end{equation}
and that of non-oscillatory motions is
\begin{equation}\label{eq:sphere:P2:thin}
P_{2}(p_0/\gamma)=1-P_{1}(p_0/\gamma)=\int_{0}^{\frac{p_{0}}{\gamma}\frac{\rho A^2}{0.7414}}g(u;\rho_{1},\rho_{2})du.
\end{equation}

For $\rho=1$, $A=1$, and $\widetilde{\mu}=\mu=\mu_1+\mu_{2}$ drawn from the Gamma distribution with $\rho_{1}=405$ and $\rho_{2}=0.01$, the probability distributions given by \eqref{eq:sphere:P1:thin}-\eqref{eq:sphere:P2:thin} are shown in Figure~\ref{fig:intpdfs-thinshell} (blue lines for $P_{1}$ and red lines for $P_{2}$). For the deterministic thin-walled tube, the critical value $p_{0}/\gamma=0.7414\underline{\mu}=3.0027$ strictly separates the cases of oscillations occurring or not. However, in the stochastic case, the two cases compete.

\section{Conclusion}\label{sec:conclude}

We provided here a synthesis on the analysis of finite amplitude oscillations resulting from dynamic finite deformations of given isotropic incompressible nonlinear hyperelastic solids, and extended this to non-deterministic oscillatory motions of stochastic isotropic incompressible hyperelastic solids with similar geometries. Specifically, we treated in a unified manner the generalised shear motion of a cuboid and the radial motion of inflated cylindrical and spherical shells of stochastic neo-Hookean or Mooney-Rivlin material. For these finite dynamic problems, attention was given to the periodic motion and the time-dependent stresses, while taking into account the stochastic model parameters, which are random variables described by given probability laws. We found that, in this case, the amplitude and period of the oscillation of the stochastic bodies are also characterised by probability distributions, and, for cylindrical tubes and spherical shells, when an impulse surface traction is applied, there is a parameter interval where both the oscillatory and non-oscillatory motions can occur with a given probability. This is in contrast to the deterministic problem where a single critical parameter value strictly separates the cases where oscillations can or cannot occur. 

The finite dynamic analysis presented here can be extended (albeit numerically) to other stochastic homogeneous hyperelastic materials (for example, using the stochastic strain-energy functions derived from experimental data in \cite{Mihai:2018:MWG}), or to inhomogeneous incompressible bodies similar to those considered deterministically in \cite{Ertepinar:1976:EA}. For incompressible bodies with inhomogeneous material parameters, the constitutive parameters of the stochastic hyperelastic models can be treated as random fields, as described in \cite{Staber:2018:SG,Staber:2019:SGSMI}. Clearly, the combination of knowledge from elasticity, statistics, and probability theories offers a richer set of tools compared to the elastic framework alone, and would logically open the way to further considerations of this type. However, as the role of stochastic effects on instabilities in finite strain elastodynamics is still in its infancy, it is important to consider the homogeneous case in the first instance.
	
If the material is compressible (unconstrained), then the theorem on quasi-equilibrated dynamics recalled by us in Section~\ref{sec:prerequisites}, is not applicable \cite[p.~209]{TruesdellNoll:2004}. As we relied on the notion of quasi-equilibrated motion to derive our analytical results for incompressible cylindrical tubes and spherical shells, the same approach cannot be used for the compressible case. Nevertheless, as seen from the generalised shear motion of a cuboid, presented in Section~\ref{sec:cubes}, more general elastodynamic problems can still be formulated where the motion is not quasi-equilibrated. However, while stochastic versions of compressible hyperelastic materials can also be obtained, as shown in \cite{Staber:2016:SG}, few theoretical results are available on the oscillatory motion of finitely deformed compressible hyperelastic solids (see, e.g., \cite{Akyuz:1998:AE}).

The analysis presented here is timely not only because ``Today, it is well understood that as soon as the probability theory can be used, then the probabilistic approach of uncertainties is certainly the most powerful, efficient and effective tool for modeling and for solving direct and inverse problems'' \cite{Soize:2013}, but also because time-dependent finite elastic deformations, although relevant to the modelling of various physical systems, have seldom been considered in more recent studies, which focused primarily on static elastic deformations or on dynamic viscoelasticity problems. Clearly, further numerical and experimental investigations of oscillatory finite deformations could help to bridge the gap between these popular areas and add some valuable insight into specific applications as well.

\appendix
\section{Additional detailed calculations}\label{sec:append}

In this Appendix, for the stochastic cylindrical and spherical shells discussed in Sections~\ref{sec:tube} and \ref{sec:sphere}, respectively, we provide detailed derivations of the general functions $G(x,\gamma)$, defined by \eqref{eq:tube:G:MR}, and $H(x,\gamma)$, defined by \eqref{eq:sphere:H:NH}, and calculate the limits of these functions in the particular cases of thin-walled and infinitely thick-walled shells.
\begin{itemize}
\item[(I)] For a Mooney-type model, the function $G(x,\gamma)$ is defined by \eqref{eq:tube:G}, where $\widetilde{\mu}=\mu_{1}+\mu_{2}\alpha^2$t. In this case, we obtain
\[
\begin{split}
G(x,\gamma)&=\frac{1}{\rho A^2}\int_{1/\sqrt{\alpha}}^{x}\left(\zeta\int_{\frac{\zeta^2+\frac{\gamma}{\alpha}}{1+\gamma}}^{\zeta^2}\widetilde{\mu}\frac{1+\alpha u}{\alpha^2u^2}du\right)d\zeta\\
&=\frac{\widetilde{\mu}}{\rho A^2}\int_{1/\sqrt{\alpha}}^{x}\left(\zeta\int_{\frac{\zeta^2+\frac{\gamma}{\alpha}}{1+\gamma}}^{\zeta^2}\frac{1+\alpha u}{\alpha^2u^2}du\right)d\zeta\\
&=\frac{\widetilde{\mu}}{\rho A^2}\int_{1/\sqrt{\alpha}}^{x}\left\{\frac{1}{\alpha^2}\left[(1+\gamma)\frac{\zeta}{\zeta^2+\frac{\gamma}{\alpha}}-\frac{1}{\zeta}\right]+\frac{1}{\alpha}\left[\zeta\log\zeta^2-\zeta\log\frac{\zeta^2+\frac{\gamma}{\alpha}}{1+\gamma}\right]\right\}d\zeta\\
&=\frac{\widetilde{\mu}}{2\alpha\rho A^2}\left(\frac{1+\gamma}{\alpha}\log\frac{x^2+\frac{\gamma}{\alpha}}{\frac{1}{\alpha}+\frac{\gamma}{\alpha }}-\frac{1}{\alpha}\log x^2+\frac{1}{\alpha}\log\frac{1}{\alpha}\right)\\
&+\frac{\widetilde{\mu}}{2\alpha\rho A^2}\left(x^2\log x^2-x^2-\frac{1}{\alpha}\log\frac{1}{\alpha}+\frac{1}{\alpha}\right)\\
&-\frac{\widetilde{\mu}}{2\alpha\rho A^2}\left(x^2\log\frac{x^2+\frac{\gamma}{\alpha}}{1+\gamma}-x^2+\frac{\gamma}{\alpha}\log\frac{x^2+\frac{\gamma}{\alpha}}{\frac{1}{\alpha}+\frac{\gamma}{\alpha}}-\frac{1}{\alpha}\log\frac{1}{\alpha}+\frac{1}{\alpha}\right)\\
&=\frac{\widetilde{\mu}}{2\alpha\rho A^2}\left(x^2-\frac{1}{\alpha}\right)\log\frac{1+\gamma}{1+\frac{\gamma}{\alpha x^2}}.
\end{split}
\]
For the thin-walled tube  \cite{Knowles:1962,Shahinpoor:1971:SN}, $\alpha=1$ and $0<\gamma\ll1$, and approximating $\log(1+\gamma)$ by $\gamma$ and $\log\left[1+\gamma/\left(\alpha x^2\right)\right]$ by $\gamma/\left(\alpha x^2\right)$, we find
\[
G(x,\gamma)=\gamma\frac{\widetilde{\mu}}{2\rho A^2}\left(x^2-1\right)\left(1-\frac{1}{x^2}\right).
\]	
For the cylindrical cavity \cite{Shahinpoor:1973}, $\gamma\to\infty$, hence
\[
G(x,\gamma)=\frac{\widetilde{\mu}}{2\alpha\rho A^2}\left(x^2-\frac{1}{\alpha}\right)\log\left(\alpha x^2\right).
\]	
\item[(II)] For a neo-Hookean-type model, the function $H(x,\gamma)$ is defined by \eqref{eq:sphere:H}, where $\widetilde{\mu}=\mu$. Following \cite{Knowles:1965:KJ}, we set the corresponding strain-energy density in the form
\[
W_{0}(u)=\frac{\mu}{2}\left(u^{-4/3}+2u^{2/3}-3\right),
\]
and denote by $W_{0}'(u)$ its first derivative with respect to $u$. Then, by standard calculations (involving integration by parts and change of variables), we obtain
\[
\begin{split}
H(x,\gamma)&=\frac{4}{3\rho A^2}\int_{1}^{x}\left(\zeta^2\int_{\frac{\zeta^3+\gamma}{1+\gamma}}^{\zeta^3}\widetilde{\mu}\frac{1+u}{u^{7/3}}du\right)d\zeta\\
&=\frac{2}{\rho A^2}\int_{1}^{x}\left(\zeta^2\int_{\frac{\zeta^3+\gamma}{1+\gamma}}^{\zeta^3}\frac{W_{0}'(u)}{u-1}du\right)d\zeta\\
&=\frac{2}{\rho A^2}\int_{1}^{x}\left\{\zeta^2\left[\frac{W_{0}(\zeta^3)}{\zeta^3-1}-\frac{W_{0}\left(\frac{\zeta^3+\gamma}{1+\gamma}\right)}{\frac{\zeta^3+\gamma}{1+\gamma}-1}+\int_{\frac{\zeta^3+\gamma}{1+\gamma}}^{\zeta^3}\frac{W_{0}(u)}{(u-1)^2}du\right]\right\}d\zeta\\
&=\frac{2}{\rho A^2}\left\{\int_{1}^{x}\zeta^2\left[\frac{W_{0}(\zeta^3)}{\zeta^3-1}-\frac{W_{0}\left(\frac{\zeta^3+\gamma}{1+\gamma}\right)}{\frac{\zeta^3+\gamma}{1+\gamma}-1}\right]d\zeta
+\int_{1}^{x}\left[\zeta^2\int_{\frac{\zeta^3+\gamma}{1+\gamma}}^{\zeta^3}\frac{W_{0}(u)}{(u-1)^2}du\right]d\zeta\right\}\\
&=\frac{2}{3\rho A^2}\left[\int_{1}^{x^3}\frac{W_{0}(u)}{u-1}du+x^3\int_{1}^{x^3}\frac{W_{0}(u)}{(u-1)^2}du-\int_{1}^{x^3}\frac{uW_{0}(u)}{(u-1)^2}du\right]\\
&+\frac{2}{3\rho A^2}\left[\int_{\frac{x^3+\gamma}{1+\gamma}}^{1}(1+\gamma)\frac{W_{0}(u)}{u-1}du+x^3\int_{\frac{x^3+\gamma}{1+\gamma}}^{1}\frac{W_{0}(u)}{(u-1)^2}du-\int_{\frac{x^3+\gamma}{1+\gamma}}^{1}\frac{\left[u(1+\gamma)-\gamma\right]W_{0}(u)}{(u-1)^2}du\right]\\
&=\frac{2}{3\rho A^2}\left[x^3\int_{1}^{x^3}\frac{W_{0}(u)}{(u-1)^2}du-\int_{1}^{x^3}\frac{W_{0}(u)}{(u-1)^2}du\right]\\
&+\frac{2}{3\rho A^2}\left[x^3\int_{\frac{x^3+\gamma}{1+\gamma}}^{1}\frac{W_{0}(u)}{(u-1)^2}du-\int_{\frac{x^3+\gamma}{1+\gamma}}^{1}(1+\gamma)\frac{W_{0}(u)}{(u-1)^2}du+\int_{\frac{x^3+\gamma}{1+\gamma}}^{1}\gamma\frac{W_{0}(u)}{(u-1)^2}du\right]\\
&=\frac{2}{3\rho A^2}\left(x^3-1\right)\int_{\frac{x^3+\gamma}{1+\gamma}}^{x^3}\frac{W_{0}(u)}{(u-1)^2}du\\
&=\frac{\mu}{3\rho A^2}\left(x^3-1\right)\int_{\frac{x^3+\gamma}{1+\gamma}}^{x^3}\frac{2u^{4/3}+4u+3u^{2/3}+2u^{1/3}+1}{u^{2/3}(u+u^{2/3}+u^{1/3})^2}du\\
&=\frac{\mu}{\rho A^2}\left(x^3-1\right)\left[\frac{2x^3-1}{x^3+x^2+x}
-\frac{2\frac{x^3+\gamma}{1+\gamma}-1}{\frac{x^3+\gamma}{1+\gamma}+\left(\frac{x^3+\gamma}{1+\gamma}\right)^{2/3}+\left(\frac{x^3+\gamma}{1+\gamma}\right)^{1/3}}\right].
\end{split}
\]
For the thin-walled shell \cite{Beatty:2011,Verron:1999:VKDR,Wang:1965}, $0<\gamma\ll1$, and
\[
\begin{split}
H(x,\gamma)&=\gamma\frac{4\mu}{3\rho A^2}\int_{1}^{x}\frac{u^6-1}{u^5}du\\
&=\gamma\frac{\mu}{\rho A^2}\frac{\left(x+1\right)\left(2x^4-x^2-1\right)}{x^3\left(x^3+x^2+x\right)}.
\end{split}
\]	
For the spherical cavity \cite{Balakrishnan:1978:BS,Knowles:1965:KJ}, $\gamma\to\infty$, hence
\[
H(x,\gamma)=\frac{\mu}{3\rho A^2}\left(x^3-1\right)\frac{5x^3-x^2-x-3}{x^3+x^2+x}.
\]	
\end{itemize}

\paragraph{Acknowledgement.} The support by the Engineering and Physical Sciences Research Council of Great Britain under research grants EP/R020205/1 for Alain Goriely and EP/S028870/1 for L. Angela Mihai is gratefully acknowledged.


\end{document}